\documentclass[fleqn,usenatbib]{mnras}

\usepackage{newtxtext,newtxmath}

\usepackage[T1]{fontenc}
\usepackage{ae,aecompl}

\usepackage{graphicx}	
\usepackage{amsmath}	
\usepackage{amssymb}	

\usepackage{graphicx}	
\usepackage{amsmath}	
\usepackage{amssymb}	
\usepackage{ltxtable}
\usepackage{pdflscape}
\usepackage{placeins}
\usepackage{longtable}
\usepackage{tabularx} 
\usepackage{tablefootnote}
\usepackage{supertabular}
\usepackage{pifont}
\newcommand{\hii}{H~{\sc ii}}
\newcommand{\ha}{H\,$\alpha$} 
\newcommand{\hb}{H\,$\beta$}
\newcommand{\hc}{H\,$\gamma$}

\newcommand{\helium}{He\,{\sc i}}
\newcommand{\heliumb}{He\,{\sc ii}}

\newcommand{\nitrogen}{[N\,{\sc ii}]}

\newcommand{\nitrogena}{[N\,{\sc i}]}
\newcommand{\oxygeniii}{[O\,{\sc iii}]}

\newcommand{\oxygeni}{[O\,{\sc i}]}
\newcommand{\oxygenii}{[O\,{\sc ii}]}

\newcommand{\sulfurt}{[S\,{\sc ii}]}

\newcommand{\ironiii}{[Fe\,{\sc iii}]}

\newcommand{\degree}{$^{\circ}$}
\def\vhel{\ifmmode{V_{{\rm HEL}}}\else{$V_{{\rm HEL}}$}\fi}
\def\vsys{\ifmmode{V_{\rm sys}}\else{$V_{\rm sys}$}\fi}
\def\kms{\ifmmode{~{\rm km\,s}^{-1}}\else{~km~s$^{-1}$}\fi}
\def\vlsr{\ifmmode{v_{\rm lsr}}\else{$v_{\rm lsr}$}\fi}

\title[VLT VIMOS integral field spectroscopy of Abell~14]{Exploring the differences of integrated and spatially resolved analysis using integral field unit data: The case of Abell~14}

\author[S. Akras et al.]{Stavros Akras$^{1,2,3}$\thanks{e-mail: stavrosakras@gmail.com},
Hektor Monteiro$^{4}$, Isabel Aleman$^{4}$, Marcos A. F. Farias$^{4}$,
\newauthor  Daniel May$^{5}$, Claudio B. Pereira$^{3}$
\\
$^{1}$ Instituto de Matem\'{a}tica, Estat\'{i}stica e F\'{i}sica, Universidade Federal do Rio Grande, Av. Italia km 8, 96203-900, Rio Grande, Brazil\\
$^{2}$ Observat\'orio do Valongo, Universidade Federal do Rio de Janeiro, Ladeira Pedro Antonio 43, 20080-090, Rio de Janeiro, Brazil\\
$^{3}$ Observat\'orio Nacional/MCTIC, Rua Gen. Jos\'{e} Cristino, 77, 20921-400, Rio de Janeiro, Brazil\\
$^{4}$ Universidade Federal de Itajub\'a, Instituto de F\'isica e Qu\'imica, Av. BPS 1303 Pinheirinho, 37500-903 Itajub\'{a}, MG, Brazil\\
$^{5}$ Instituto de Astronomia, Geof\'{i}sica e Ci\^{e}ncias Atmosf\'{e}ricas, Universidade de S\~{a}o Paulo, 05508-090, S\~{a}o Paulo, Brazil
}
\date{Accepted XXX. Received YYY; in original form ZZZ}

\pubyear{2015}

\begin{document}
\label{firstpage}
\pagerange{\pageref{firstpage}--\pageref{lastpage}}
\maketitle

\begin{abstract}
We present a new approach to study planetary nebulae using integral field spectroscopy. VLT@VIMOS datacube of the planetary nebula Abell~14 is analysed in three different ways by extracting: (i) the integrated spectrum, (ii) 1-dimensional simulated long slit spectra for different position angles and (iii) spaxel-by-spaxel spectra. These data are used to built emission-line diagnostic diagrams and explore the ionization structure and excitation mechanisms  combining data from 1- and 3- dimensional photoionization models. The integrated and 1D simulated spectra are suitable for developing diagnostic diagrams, while the spaxel spectra can lead to misinterpretation of the observations.
We find that the emission-line ratios of Abell~14 are consistent with UV photo-ionized emission; however, there are some pieces of evidence of an additional thermal mechanism. The chemical abundances confirm its previous classification as a Type I planetary nebula, without spatial variation. We find, though, variation in the ionization correction factors (ICFs) as a function of the slit position angle. The star at the geometric centre of Abell~14 has an A5 spectral type with an effective temperature of $T_{\rm eff}$~=~7909$\pm$135~K and surface gravity log($g$)~=~1.4$\pm$0.1~cm~s$^{-2}$. Hence, this star cannot be responsible for the ionization state of the nebula. {\it Gaia} parallaxes of this star yield distances between 3.6 and 4.5~kpc in good agreement with the distance derived from a 3-dimensional photoionization modelling of Abell~14, indicating the presence of a binary system at the centre of the planetary nebula.

\end{abstract}




\section{Introduction}

\begin{figure}
    \centering
    \includegraphics[scale=0.55]{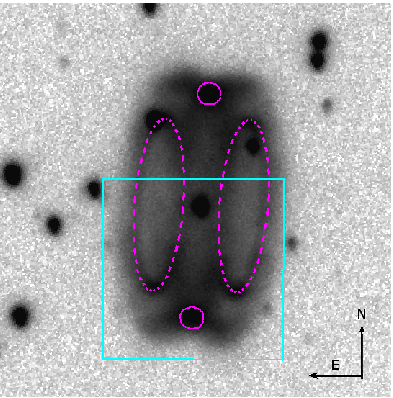}
    \caption{\nitrogen~$\lambda$6583 image of Abell~14 obtained with the Aristarchos telescope at the Helmos Observatory in Greece. The field of view of the images is 60$\times$60~arcmin. The western and eastern ring-like structures are indicated with the two magenta dashed-line ellipses. The bright northern and southern spots in the equatorial region are indicated with the magenta circles. The cyan box illustrates the southern field (27$\times$27~arcsec) of the nebula observed with VIMOS at VLT. }
    \label{fig:abell14new}
\end{figure}

Integral field spectroscopy (IFS) provides unique information of extended sources. The disadvantages of the first-generation integral field units (IFUs) such as the small field of view and low spectral resolution have been overcome by the new generations of instruments developed over the past few yr either in the optical regime (e.g. MUSE, MEGARA, WiFeS, and SITELLE)\footnote{Multi Unit Spectroscopic Explorer \citep[(MUSE; ][]{Bacon2010}, Multi-Espectr\'ografo en GTC de Alta Resoluci\'on para Astronom\'ia \citep[MEGARA;,][]{GildePaz2016}, Wide-Field Spectrograph \citep[WiFeS; ][]{Dopita2007}, Spectro-Imageur \`a Trasform\'ee de Fourier pour l'\'{E}tude en Long et en Large 
desraies d'\'Emission \citep[SITELLE; ][]{Grandmont2012}} or in the infrared regime (e.g. SIFONI, NIFS, and FRIDA)\footnote{Spectrograph for INtegral Field Observations in the Near Infrared \citep[SINFONI; ][]{Eisenhauer2003,Bonnet2004},  Near-Infrared Integral Field Spectrometer \citep[NIFS; ][]{McGregor2003}, inFRared Imager and Dissector for the Adaptive optics system of the Gran Telescopio Canarias \citep[FRIDA; ][]{Cuevas2008}.}. All these IFUs have been widely used to study galaxies and emission-line objects such as \hii~regions and planetary nebulae (PNe). The Calar Alto Legacy Integral Field Area Survey \citep[CALIFA;][]{sanchez2012} corroborates the power of IFS in the study of extended galaxies. 

IFU data have made possible the development of more sophisticated emission-line diagnostic diagrams (DDs) and the investigations of the mechanisms responsible for the spectral emission-lines in  both spatial directions. Different regions across galaxies have been found to display different emission-line ratios (e.g. LINERs-like, Seyfert-like, star formation-like) and occupy distinct regimes on the Baldwin, Phillips \& Terlevich \citep[BPT,][]{Baldwin1981} DDs \cite[e.g.][]{singh2013,Ho2014,Belfiore2016}. The same methodology has also been applied to \hii~regions \citep{sanchez2013}. However, only a few PNe had been observed and studied with IFS, the majority of them during the last 5-6~yr \cite[e.g.][]{Tsamis2008,Danehkar2013,Hektor2013,Danehkar2014,Ali2015,Ali2016,Danehkar2015b,Danehkar2015,Basurah2016,Danehkar2016,Walsh2016,Ali2017,Dopita2017,Dopita2018,Walsh2018,Ali2019} and imaging fourier transform spectroscopy \citep{Lagrois2015}.

In contrast to IFS, the position of the slit(s) in long-slit spectroscopy is usually selected based on the morphology of the nebula and/or on the position of distinct features, such as low-ionization structures \citep[LISs: knots, jets, and filaments; e.g.,][]{Balick1993,Balick1994,akrasLIS2016,Derlopa2019} or bipolar outflows \citep[e.g.][]{Guerrero2013,Akras2015,Ramos2018}.

In this work, we present the first VLT@VIMOS (VIsible Multi-Object Spectrograph) IFU data of Abell~14, an intriguing nebula for which strong \oxygeni~$\lambda$6300 \citep[][]{Henry2010} and \nitrogena~$\lambda$5200 \citep[][]{Bohigas2003} emission-lines have been reported. The detection of only one of these lines is confirmed in each study above. Such discrepancy may be related to the slit position in each observation. Based on 3D photoionization modelling, \citet{akras2016} argued for an additional thermal mechanism, probably shock interactions, to explain the observed intensities of these two emission-lines. All these make Abell~14 an ideal object to be studied through IFS and explore the distribution of emission-lines in two spatial dimensions as well as its BTP DDs in a spatially resolved manner.

The paper is organized as follows. The VLT observations and data reduction are presented in Section~2. In Section~3, we describe the different approaches used in this work to analyse the VIMOS datacubes: integrated spectrum, simulated 1D long-slit spectroscopy and spaxel-by-spaxel analysis. In Section~4, we present a number of emission-line DDs generated using the fluxes obtained from the three aforementioned methods, as well as emission-line fluxes from a grid of {\sc cloudy} \citep[][]{Ferland2017} models obtained from the Mexican Million Models data base \citep[3MdB; ][]{Morisset2015}. In this section, we also discuss how chemical abundances and ICFs vary as function of the slit position. The optical spectrum of the central star of Abell~14 is also presented and we  discuss the possibility of being a binary system. We summarize our conclusions in Section 5.

\section{Observations and data reduction}

\begin{table*}
\caption[]{VLT@VIMOS IFU observing log}
\label{table1}
\begin{tabular}{lllllll}
\hline 
Name & Date & Grism & Exp. time     &  RA.       & Dec.  \\     			
     &      &       & (s)           &  (J2000.0) & (J2000.0) \\   
\hline 
HRblue                 & 28/01/2017 & HR blue     & 2x1000 & 06:11:08.7 & 11:46:34.2\\
HRorange               & 21/02/2017 & HR orange   & 2x1000 & 06:11:08.7 & 11:46:34.2\\
\hline 
Sky HRblue             & 28/01/2017 & HR blue     & 1x300  & 06:11:05.7 & 11:46:20.1 \\
Sky HRorange           & 21/02/2017 & HR orange   & 1x300  & 06:11:05.7 & 11:46:20.1\\
\hline
LTT~3864$^{\dag}$      & 28/01/2017 &  HR blue    &  1x128 & 10 32 13.6 & 10 32 13.6  \\   
CD~-32 9927$^{\dag}$   & 21/02/2017 &  HR orange  &  1x29  & 14 11 46.3 & -33 03 14.3\\
\hline
\end{tabular} 
\begin{flushleft}
$^{\dag}$ Sky calibrations sources.
\end{flushleft}
\end{table*}

Abell~14 was observed at the VLT of ESO (UT3 Melipa) using the IFU mode of the VIMOS instrument \citep{Lafevre2003}. The data were obtained on 2017 January 28 and February 21 [ID: 098.D-0436(A); PI: S. Akras] in service mode.

The high resolution blue grism (HRblue) with a dispersion of 0.571~\AA\ pixel$^{-1}$ and a spectral resolution of 1440 was used to cover the spectral range from 3700 to 5220~\AA. The orange grism (HRorange) and the blocking filter GG435 with dispersion of 0.6~\AA\ pixel$^{-1}$ were used to cover the spectral range from 5250 to 7400~\AA. The field of view of 27 $\times$ 27~arcsec with a spatial resolution of 0.67~arcsec~pixel$^{-1}$ and 40 $\times$ 40 fibres (i.e. spaxels), available for these two high resolution grisms, was selected. Such a configuration was chosen to allow the observation of the entire Abell~14 \citep[41 $\times$ 23~arcsec; ][]{akras2016} with two pointings (north and south). 

Because of the presence of two field stars at the northern part of the nebula (see Figure~\ref{fig:abell14new}), the priority was given to the southern part, which is free from field stars. Due to bad weather conditions, only the southern part of Abell~14 was eventually observed. Abell~14 is a highly symmetrical and homogeneous nebula with negligible flux variations, as it can be seen in Figure~\ref{fig:abell14new} \citep[see also ][]{akras2016}. This symmetry assures that our results and conclusions obtained for the southern half of the nebula can also be applied to the other half. The ring-like structures in the eastern and western directions of the nebula as well as the bright-spots in the equatorial regions (along the north-south direction) can easily be seen in the \nitrogen~$\lambda$6583 line image. 

To increase the signal-to-noise (S/N) ratio of emission-lines, two exposures of 1000~s per grism were obtained as well as one additional exposure of 300~s in an area close to Abell~14 in order to subtract the background skylines from the emission-line images. The seeing during the observations was varying between 0.7 and 1.3 arcsec. For the flux calibration of the data, two standard stars were observed, one for each grism. In Table~\ref{table1}, we present the observing log. 

The reduction of the data was carried out with the VIMOS pipelines available in the instrument website\footnote{http://www.eso.org/sci/facilities/paranal/instruments} and our own code to reconstruct the final datacubes and extract the emission-lines maps. 

\section{Methodology}
Originally, BPT diagnostic diagrams were first constructed using long-slit spectroscopic data \citep[][]{Baldwin1981}. It is thus coherent to explore whether emission-line ratios derived from a spaxel-by-spaxel analysis of IFU datacubes can be used to study resolved PNe and trace their excitation mechanisms through the BPT diagrams.

Three approaches were followed for the analysis of Abell~14: (1) integrated spectrum, (2) spaxel-by-spaxel (emission-line maps), and (3) 1D simulated long-slit spectra in distinct position angles (PAs) relative to the position of the central star.

The line fluxes in each spaxel (integrated spectrum or 1D simulated long-slit spectra) are corrected for interstellar extinction using the extinction coefficient c(\hb) derived from the Balmer \ha/\hb~line ratio (equation~1), from the same spaxel (the observed southern part of the nebula or the simulated slits) considering for all the cases the interstellar extinction law by \cite{Fitzpatrick1999} and R$_{\rm{v}}$=3.1. The formulae used are:

\begin{eqnarray}
 c_{i}( {\rm H}\beta)=\left(\frac{1}{0.348}\right)
  {\rm Log}\left[\frac{F_{i}( {\rm H}\alpha)/ F_{i}( {\rm H}\beta)}{2.85}\right]
\end{eqnarray}

\begin{eqnarray}
 \left[\frac{F_{i}(x)}{F_{i}( {\rm H}\beta)}\right]_{corr} = \left[\frac{F_{i}(x)}{F_{i}( {\rm H}\beta)}\right]_{obs}~10^{c_{i}( {\rm H}\beta)\times[f(\lambda)-f( {\rm H}\beta)]}
\end{eqnarray}

\noindent where $F_{i}(x)$ is the flux of the line $x$ in the spaxel $i$, $c_{i}$ is the extinction coefficient in the spaxel $i$, and $f(\lambda)$ is the extinction law. The \hc/\hb\ ratio is not used to estimate the extinction coefficient due to the very low S/N of \hc.

The integrated emission-line fluxes from the observed portion of Abell~14 are derived by summing up all the spaxels and discarding those with (i) stellar emission, (ii) low signal-to-noise fluxes, and (iii) F(\ha)$<$F(\hb)*2.85 (after reddening correction), which implies unrealistic (negative) values for the extinction coefficient c(\hb). The 1D long-slit spectra were extracted simulating slits of 2~arcsec widths in different PAs relative to the central star's position from -90 to 90\degree. The final flux was calculated by summing up the spaxels' fluxes. 

All the nebular emission-lines measured from the integrated spectrum and the 1D simulated long-slit spectra are listed in Table~\ref{table2}. The observed line fluxes from the long-slit spectra of \cite{Bohigas2003} and \cite{Henry2010} are also presented for direct comparison with the new data.

\subsection{0D approach: Integrated spectrum}
For the integrated spectrum, we simulated a slit of width and length such that the whole observed nebula was fitted in this slit. The extinction coefficient c(\hb) and emission-line ratios were calculated several times by rotating the line maps from 0 to 360\degree\ in steps of 2\degree. This exercise was performed to ensure that the rotation of line maps or of the simulated slit does not affect the resultant line fluxes. We find a c(\hb) equal to 0.46 with a standard deviation (SD) of 0.003. The very low SD confirms the negligible effect of rotation. Integrated emission-lines are also described by very low SDs. The line fluxes of the integrated spectrum (PA=0\degree) are presented in the fourth column of Table~\ref{table2}.

Due to the versatility of IFU data, the datacube was also used as input to the {\sc alfa}\footnote{Automated Line Fitting Algorithm {\sc alfa} identifies and fits the lines in a spectrum with a Gaussian profile \citep{Wesson16}. All line intensities obtained from this procedure can be seen in the third column of Table~\ref{table2}.}. The c(\hb) value is estimated to be 0.69$\pm$0.05. This value is higher than the one derived from the line maps because {\sc alfa} does not apply any criteria to the emission-line in each spaxel. 

The extinction coefficient of Abell~14 is evidently lower than the previous long-slit studies (0.98, \citealt{Bohigas2003}; 0.88, \citealt{Henry2010}). This implies a difference in the \ha\ and/or \hb\ fluxes. Nevertheless, a reasonable agreement between our \hb\ flux from the simulated slit at PA=0\degree\ and that from \cite{Bohigas2003} is found (see~Table~\ref{table2}). 

Our integrated line intensities agree better with the results obtained from \cite{Bohigas2003} than those from \cite{Henry2010}. Even the \nitrogena\ $\lambda$5200 line is consistent with Bohiga's results. On the other hand, the \oxygeni\ $\lambda$6300 line, which is detected by Henry but not in Bohigas's data, is marginally detected in our IFU data with half of the intensity measured by \cite{Henry2010} and very low S/N ratio. The \oxygeni~$\lambda$6300 line map shows no clear morphological structure which can be indicative of poor sky subtraction.

\begin{figure}
    \centering
    \includegraphics[scale=0.33]{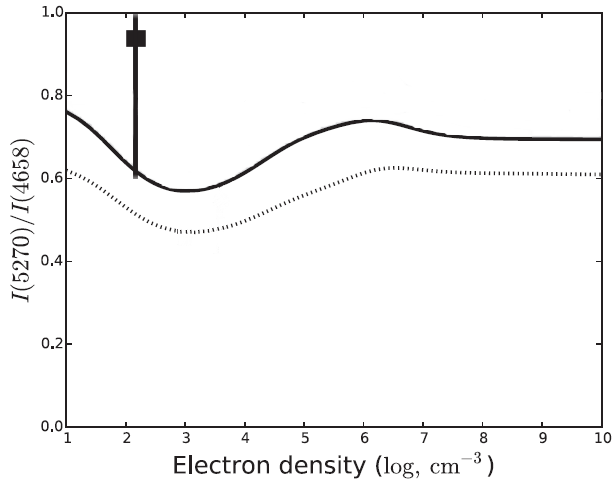}
    \caption{The theoretical \ironiii\ $\lambda\lambda$5270/4658 line ratio as a function of electronic density for two different electronic temperatures at  9000~K(solid line) and 15000~K (dotted line) derived from \citep{Laha2017}. 
    The black square with error bars indicates the value 0.95$\pm$0.33 we found for Abell~14.
    }
    \label{fig:feiii}
\end{figure}

Besides the detection of the common nebular emission-lines, two \ironiii\ lines centred at 4658~\AA\ and 5270~\AA\ are also detected very close to the noise level with values of 13.5 and 12.9 relative to \hb=100, respectively. The \ironiii\ $\lambda\lambda$5270/4658 line ratio can be used as a density-indicator \citep[see][]{Laha2017}. Its value is calculated (0.95$\pm$0.33) and it is consistent with the electronic density of 100-200~cm$^{-3}$ derived from the \sulfurt\ diagnostic lines \citep{Bohigas2003,Henry2010,akras2016} for an electronic temperature of ~10000~K (Figure~\ref{fig:feiii}). Both the lines have been detected in several PNe, for instance Mz~3, NGC~2392, NGC~6543, and NGC~7027, with X-ray emission also detected \citep{Perinotto1999,Zhang2002,Zhang2012}. 

Due to the very low S/N of the \ironiii\ and \oxygeni\ lines, this should be considered as a tentative detection and further high S/N observations should be carried out.

\begin{figure}
    \centering
    \includegraphics[width=\columnwidth]{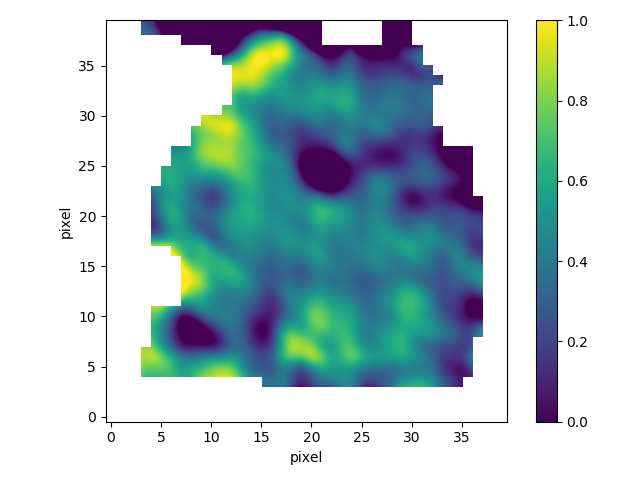}
    \caption{Map of the extinction coefficient c(\hb). The overlaid contours indicate the \nitrogen\ 6584~\AA\ intensity levels, showing the position of the rings and main structure of the nebula. North is down and east is to the right.}
    \label{fig:cmap}
\end{figure}

\begin{table*}
\caption[]{~Integrated and 1D simulated long-slit emission-line intensities for PAs=-90,-80,-20, and 0$^{\circ}$ obtained from the VIMOS datacube.} \label{table2}
\begin{tabular}{cccccccccc}

\hline
Ion & $ \lambda $ & $I \left( \lambda \right)^{\dag}$ & $I \left( \lambda \right)^{\dag\dag}$ & \multicolumn{4}{c}{PA} & B03$^a$  & H10$^b$  \\

Ion & $ ($\AA$) $ & {\sc alfa} & Maps  &  -90$^{\circ}$ & -80$^{\circ}$ & -20$^{\circ}$ & 0$^{\circ}$ &  &   \\

\hline
[O~{\sc ii}]   & 3727.00 &  486 $\pm$ 82    & 442     & 366  & 421   & 347    & 604    & -      & 326\\ 

[Ne~{\sc iii}] & 3868.75 &  95.5 $\pm$ 20.2 & 82.4    & 112  & 110   & 74.3   &  58.6  & -      & 63.5\\

H~{\sc i}      & 3970.07 &  64.7 $\pm$ 19.1 &  -      & -    & -     & -      & -      & -      & -   \\

H~{\sc i}      & 4340.47 &  37.6 $\pm$ 5.7  &  -      & -    & -     & -      & -      & 41.0   & 60.7\\

[Fe~{\sc iii}] & 4658.10 &  13.5 $\pm$ 2.0  &  -      & -    & -     & -      & -      & -      & -    \\

He~{\sc ii}    & 4685.68 &  24.0 $\pm$ 2.4  &  25.8   & 43.5 & 43.6  &  25.8  &  24.3  & 33.7   & 26.4\\

H~{\sc i}      & 4861.33 &  100 $\pm$ 3     & 100     & 100  & 100   & 100    & 100    & 100    & 100 \\

[O~{\sc iii}]  & 4958.91 &  103 $\pm$ 2     & 94.4    & 146  & 149   & 101    &  88.3  & 98.5   & 157 \\

[O~{\sc iii}]  & 5006.84 &  304 $\pm$ 3.0   & 281     & 449  & 453   & 310    & 260    & 307    & 475  \\

[N~{\sc i}]    & 5200.26 &  24.5 $\pm$ 3.0  & 36.6    & 36.5 & 33.7  &  42.8  &  23.5  & 28.8   & -   \\

[Fe~{\sc iii}] & 5270.40 &  12.9 $\pm$ 4.2  &  -      & -    & -     & -      & -      & -      & -    \\

[N~{\sc ii}]   & 5754.60 &  24.0 $\pm$ 2.7  & 26.1    & 21.5 & 20.9  & 18.4   & 22.5   & 29.0   & 40.7\\

He~{\sc i}     & 5875.66 &  17.0 $\pm$ 3.3 & 18.6     & 10.7 & 12.3  & 17.7   & 16.9   & 15.6   & 22.4\\

[O~{\sc i}]    & 6300.34 &  14.7 $\pm$ 4.1 &  -       & -    &  -    & -      & -      & -      & 30.5\\

[N~{\sc ii}]   & 6548.10 &  436 $\pm$ 16   & 422      & 301  & 316   & 420    & 439    & 479    & 552  \\

H~{\sc i}      & 6562.77 &  276 $\pm$ 9    & 284      & 282  & 283   & 286    & 285    & 280    & 286\\

[N~{\sc ii}]   & 6583.50 & 1350 $\pm$ 50   & 1298     & 956  & 989   & 1308   & 1365   & 1435   & 1650 \\

[S~{\sc ii}]   & 6716.44 &  65.4 $\pm$ 3.1 &   62.7   & 43.4 & 45.7  &  58.5  &  60.4  & 57.1   & 66.5 \\

[S~{\sc ii}]   & 6730.82 &  44.2 $\pm$ 3.4 &   44.9   & 33.6 & 31.8  & 44.1   &  42.7  & 42.9   & 48.7\\

[Ar~{\sc iii}] & 7135.80 &  21.3 $\pm$ 2.2 &   19.6   & 21.9 & 19.2  & 19.9   &  18.3  & 19.5   & 10.8 \\
 \hline
 F(\hb)$^{c}$ & $-$    & $33.5$  & 34.6  & 2.9& 3.0& 5.9 & 6.2& 5.9 & 1.51 \\
 c(\hb) & $-$  & 0.69 $\pm$ 0.05 & 0.51 & 0.54      & 0.56   & 0.43     & 0.46  & 0.98  & 0.88 \\
\hline
\end{tabular}
\begin{flushleft}
$^\dag$ Line intensities have been estimated from the whole nebula observed with VIMOS (integrated spectra) using the code {\sc alfa}\\
$^{\dag\dag}$ Line intensities have been estimated from the whole nebula observed with VIMOS (integrated spectra) using the line maps\\
$^a$ \citep{Bohigas2003}: Slit width 3~arcsec oriented at north--south direction (PA=0), $^b$ \citep{Henry2010}: Slit width 2~arcsec oriented at PA=100, $^c$ in units of 10$^{-15}$ erg~cm$^{-2}$~s$^{-1}$\end{flushleft}

\end{table*}

\subsection{1D approach: Simulated long-slit spectra}
For the 1D long-slit approach, we simulated a number of slits with 3 spaxel width (equivalent to 2~arcsec), for different PAs from -90 (eastern) to 90\degree\ (western) with a step of 2\degree. The average value of c(\hb) is 0.47 with SD=0.05 and agrees with the value derived from the integrated spectrum, but its higher SD implies some variation with PA.

\subsection{2D approach: Spaxel-by-spaxel analysis}
An analysis of the emission-line maps spaxel-by-spaxel was also performed. In this case, we calculated the c(\hb) and the emission-line fluxes for each individual spaxel. The average c(\hb) of Abell~14 is equal to 0.55 with SD=0.45. This clearly demonstrates a significant variation in the extinction coefficient from one spaxel to another. Figure~\ref{fig:cmap} displays the map of the extinction coefficient showing its spatial variation. Particularly is noticeable the large extinction in the region of the western ring.

Given that VIMOS IFU has 40$\times$40 fibres (i.e. spaxels), there are available 1600 spectra. It is, thus, very likely that some of the measurements are problematic, for instance at the border of the CCD with very low SN ratio. Therefore, it is necessary to find (if there are) those problematic values (outliers) and exclude them from the final calculations.

To find the outliers, we first calculated the upper (Q1, 25 percent) and lower (Q3, 75 percent) quartiles. Then, the outliers were found as those data points (spaxels' measurements) in the emission-line maps with values $<$Q1-1.5*(Q3-Q1) and $>$Q3+1.5*(Q3-Q1) \citep[][]{upton1997,kokoska2000}. The new value of c(\hb), excluding 5 per cent of the total number of spaxels' measurements, is found to be 0.47 with SD=0.29. This result agrees better with the value derived from the integrated spectrum and it implies that the outliers, mainly from areas with low S/N,  affect the resultant average values from spaxels. The same procedure is followed for all the emission-lines.

\section{Results}

In the following subsections, we present the 2D maps for the most representative emission-lines of Abell~14 together with a morphological analysis. Then, we discuss the excitation mechanisms of Abell~14 through the BPT DDs. These diagrams are constructed combining data from the 1D simulated long-slit VIMOS spectra, the spaxel-by-spaxel spectra, the 1D photoionization models from the Mexican Million Models data base (3MdB, \citealt{Morisset2015}) and a 3D photoionization model \citep{akras2016}. We finish with the analysis of the electronic density and temperature as well as the ionic and total chemical abundances of the nebula. 

Concerning the 3MdB, it is worth mentioning some information about the grid of photoionization models. This grid of models models was constructed by \citet{Morisset2015} using the code {\sc cloudy} \citep{Ferland2017}. Various parameters were set as free variables and varied within a specific range of values. These parameters are as follows: the effective temperature ($T_{\rm eff}$) and luminosity (L) of the central star, the hydrogen density, the inner radius of the nebular shell, the metallicity, and the chemical and dust composition.

Yet, not all the models of the grid depict real PNe and a number of criteria were considered in order to get a more representative grid of models \citep{Inglada2014}. These criteria are as follows: (a) hydrogen mass$<$1~M$\odot$,~(b)~$T_{\rm eff}$ and L consistent with evolutionary models, (c) 10$^{-13}<$surface brightness$<$10$^{-11}$~erg~s$^{-1}$~cm$^{-2}$~arcsec$^{-2}$, and (d) 2$\times$10$^{53}\leq n_HR_{out}^3\leq$3$\times$10$^{56}$. For more details about the grid of PNe models, we refer the interested readers to the paper from \cite{Inglada2014}. In order to assure that Abell~14 is compared with the equivalent models  (e.g. spectral energy distribution, ionization degree), we plot in the BPT diagrams only those models for which the temperature of the central star covers a range of values from 100000 to 200000~K and the nebula is 80\% matter bounded.

\subsection{Morphological analysis from line maps}

\begin{figure*}
\includegraphics[scale=0.65]{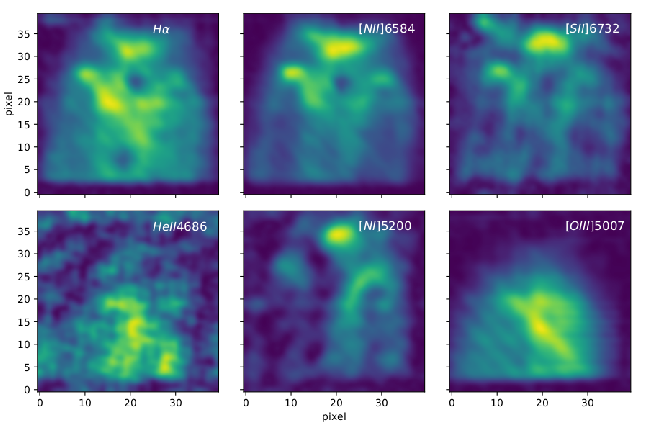}
\caption[]{2D maps of the \ha, \nitrogen\ 6584\AA, \sulfurt\ 6716,6731 \AA, \heliumb\ 4846 \AA, \nitrogena\ 5200 \AA\ and \oxygeniii\ 5007\AA\ emission-lines for the southern part of Abell~14 observed with VIMOS at VLT. North is down and east is to the right.}
\label{linemaps}
\end{figure*}

Figure~\ref{linemaps} illustrates the 2D maps of the brightest emission-lines of Abell~14, i.e., \ha, \nitrogen\ 6584~\AA, \sulfurt\ 6716,6731~\AA, \heliumb\ 4846~\AA, \nitrogena\ 5200~\AA\, and \oxygeniii\ 5007~\AA. The two ring-like structures in the eastern and western parts of the nebula, as well as the bright spot in the southern direction of the equatorial region can barely be seen in the \nitrogen\ and \sulfurt\ lines and almost disappear in the \ha\ line. The reason for that is the very low spatial resolution of the VIMOS images compared to the narrow-band images obtained from the Aristarchos telescope (see Figure~\ref{fig:abell14new}). 

The two ring-like structures appear brighter at the southern edge displaying a bright spot. Moreover, the western ring is brighter compared to the eastern one due to the orientation to the nebula. This is consistent with its recent 3D morpho-kinematic model \citep{akras2016}. The southern bright spot at the equatorial region, which is noticeable in all lines besides \oxygeniii\ and \heliumb\ may be the result of the projection of a larger fainter hourglass structure or the interaction between two larger but fainter ring-like structures. The angular size of the observed portion of the nebula in the \nitrogen\ and \sulfurt\ lines, measured from the outer edge to the central star, is 20~arcsec in the NS direction and 22~arcsec in the EW direction.

\begin{figure*}
\includegraphics[scale=0.38]{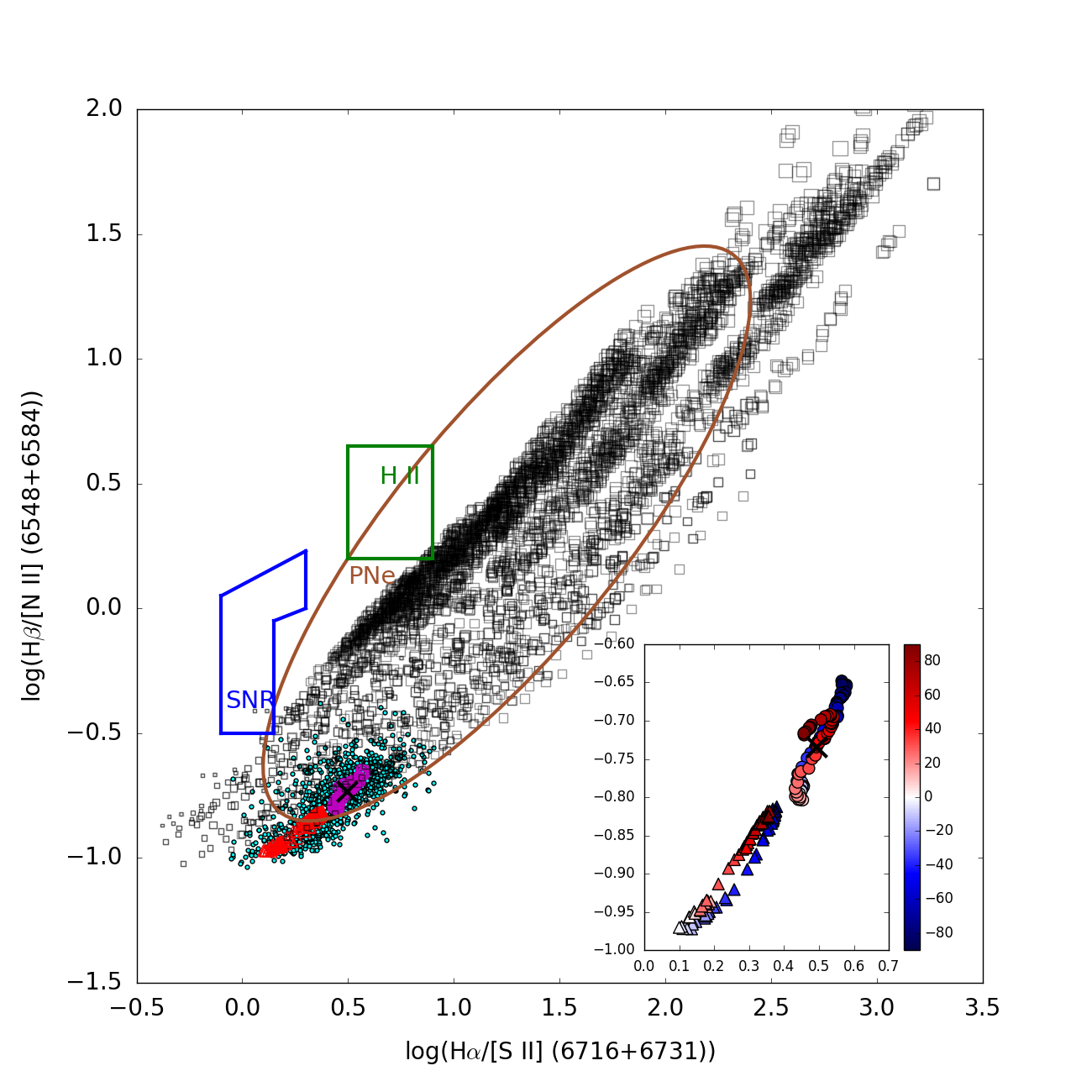}
\caption[]{log(\ha/\nitrogen) and log(\ha/\sulfurt) DD. The small cyan dots represent the spaxels' ratio, the purple circles and red triangles correspond to the 1D simulated long-slit ratios from the VIMOS observations and the {\sc mocassin} \cite[][]{Ercolano2003,Ercolano2005} model, respectively of Abell~14 constructed by \cite{akras2016} for different PAs. The open black squares represent 1D photoionization {\sc cloudy} \citep[][]{Ferland2017} models from the Mexican Million Models data base \citep[3MdB,][]{Morisset2015}. The size of the open black squares indicates the logU parameters for each model. The bigger the symbol, the higher the logU. The areas occupied by PNe, \hii~regions and supernova remnants are also marked. The inner plot illustrates the distribution of the 1D simulated long-slit ratios. The colour bar indicates the PA of the slit relative to the central star's position }
\label{DD1}
\end{figure*}

Regarding the high-excitation lines such as the \heliumb\ and \oxygeniii\ lines, none of the ring-like structures or the southern bright spot are detected. The nebula displays a more ellipsoidal structure concentrated in the centre. The size of the observed part of the nebula in the \oxygeniii\ line is 15x20~arcsec. In the \nitrogena\ 5200~\AA\ emission-line map, Abell~14 displays a diffuse emission in the eastern ring as well as the southern bright spot. Shock interactions may be responsible for the strong \nitrogena\ 5200~\AA\ emission in these particular regions of the nebula. 

Hydrodynamic simulations have shown that the interaction of jets or collimated outflows in the polar direction, due to the presence of a binary system, with a dense shell can explain the formation of barrel-like PNe \citep{akashi2018}. The hydrodynamic simulations R1, R3 and R5 resemble the morphology of Abell~14 \citep[see Figure 10 in][]{akashi2018}. In particular, the R1 model reproduces the bright spots at each ring-like structure observed in Abell~14. Surprisingly, the viewing angle of the R1 model (20\degree) matches with the inclination angle of 22\degree$\pm$4\degree\ in the east-west direction obtained from its 3D morpho-kinematic model \citep{akras2016}.

\begin{figure*}
\includegraphics[scale=0.38]{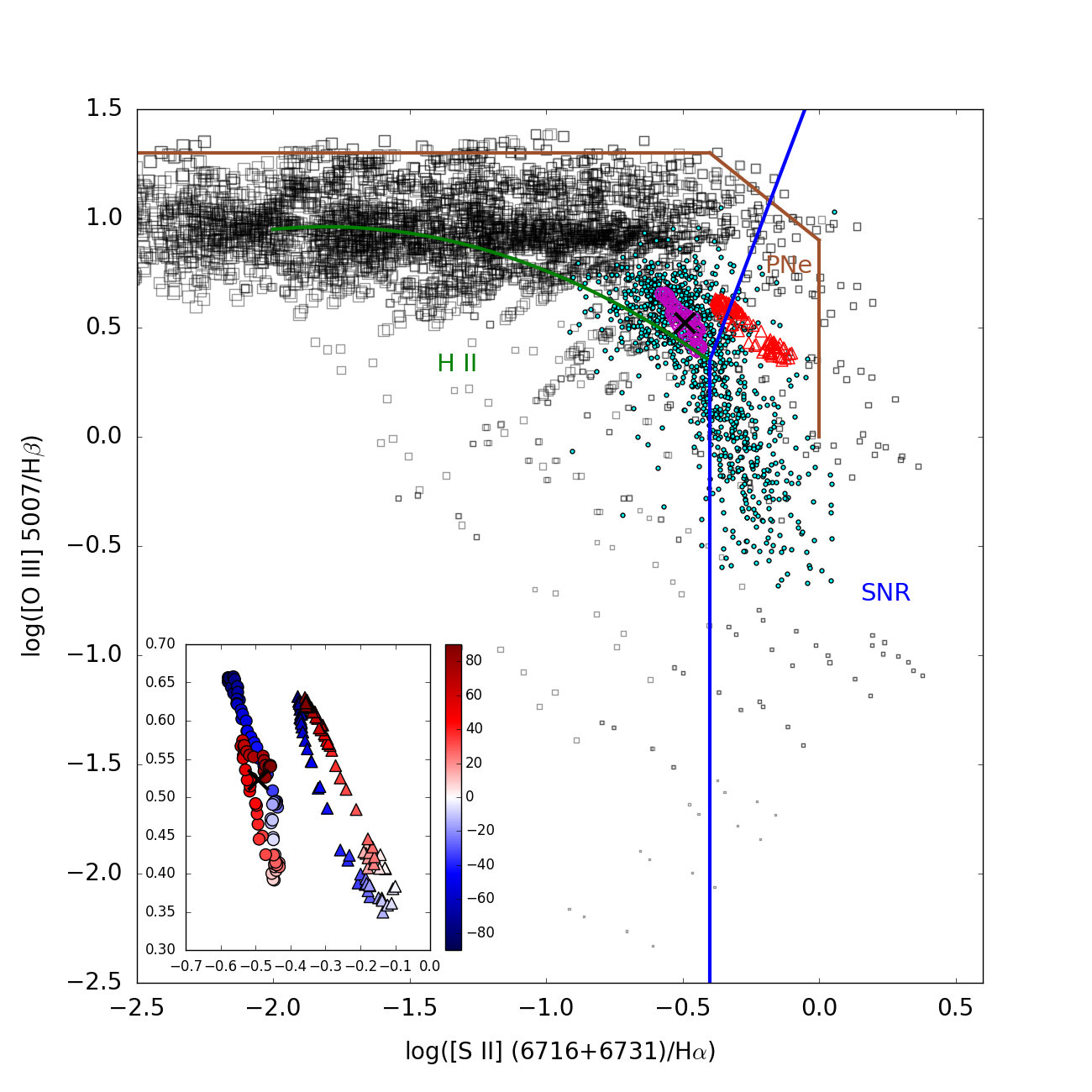}
\caption[]{The same as in Figure~\ref{DD1} for log(\sulfurt/\ha) versus log(\oxygeniii~5007/\hb).}
\label{DD2}
\end{figure*}

\begin{figure*}
\includegraphics[scale=0.38]{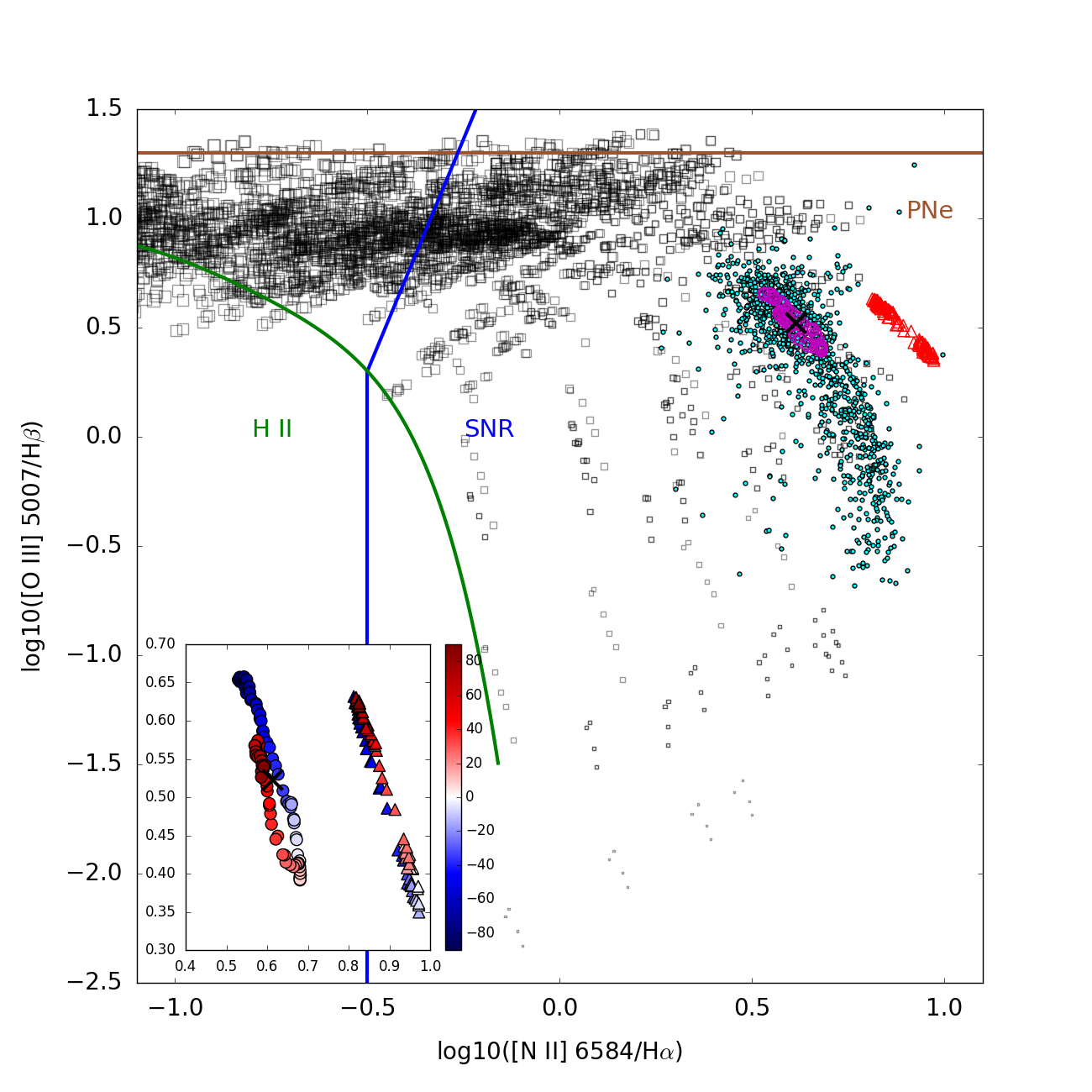}
\caption[]{The same as in Figure~\ref{DD1} for log(\nitrogen~6584/\ha) versus log(\oxygeniii~5007/\hb).}
\label{DD3}
\end{figure*}

\begin{figure*}
\includegraphics[scale=0.38]{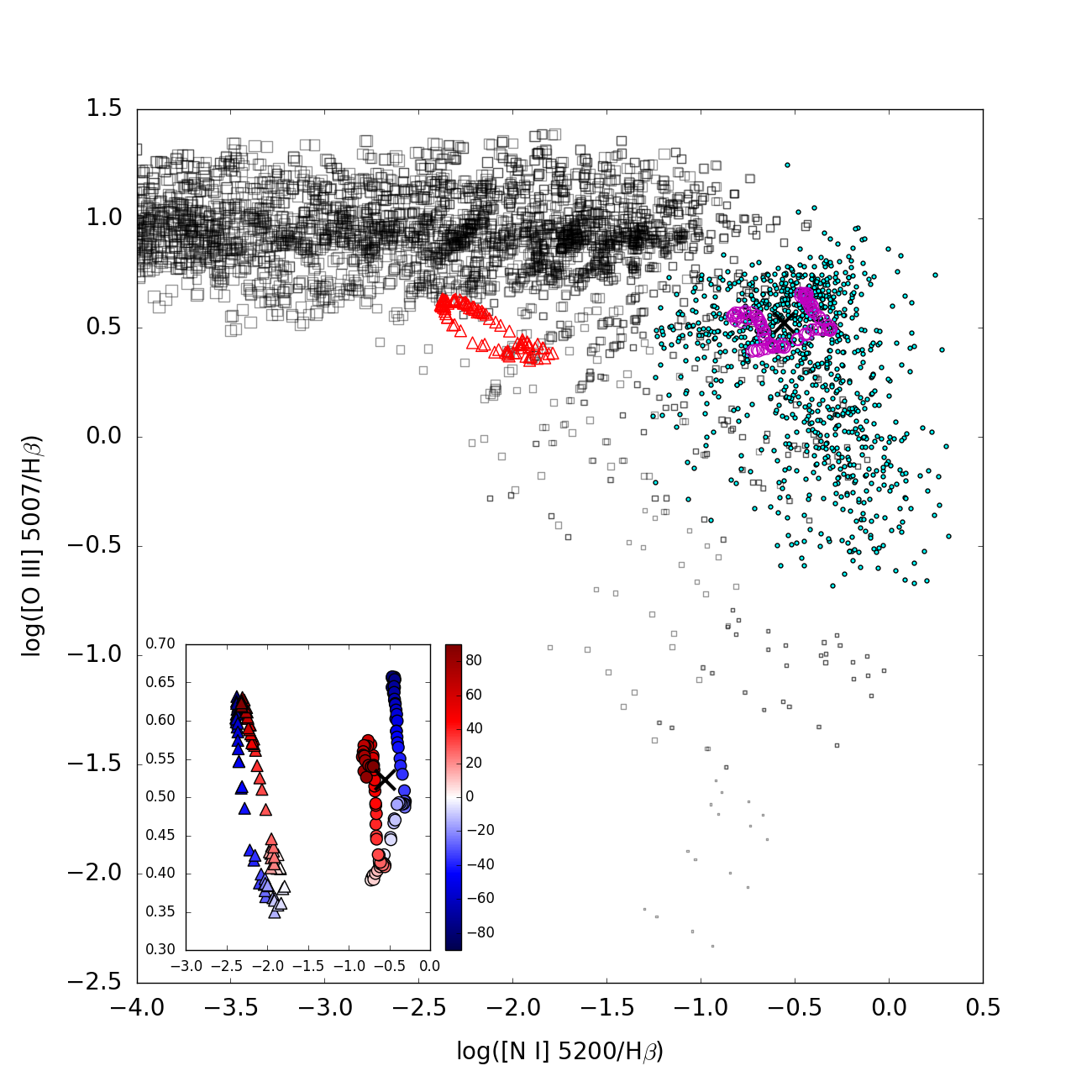}
\caption[]{The same as in Figure~\ref{DD1} for log(\nitrogena~5200/\hb) versus log(\oxygeniii~5007/\hb).}
\label{DD4}
\end{figure*}

\subsection{Emission-line diagnostic diagrams}

First \cite{Baldwin1981} and later \cite{Veilleux1987} proposed a number of DDs for galaxy classification based on the \oxygeniii/\hb, \nitrogen/\ha, \sulfurt/\ha, and \oxygeni/\ha\ line ratios, known as BPT or VO diagrams. The physical separation between star-forming, Seyfert and LINER galaxies on the emission-line ratios is attributed to the ionization mechanism. Shock with velocities higher than 100~\kms\ can provide a reasonable explanation for the strong emission of low-ionization lines and the position of LINER on the BPT diagrams \citep[e.g. ][]{Ho2014}.

Despite all the effort to distinguish different type of galaxies and explore the ionization structure, only few studies have been done in PNe. The most common and widely known DD is between log(\ha/\nitrogen) and log(\ha/\sulfurt) line ratios proposed by \cite{SMB1977} (SMB). A statistical analysis of this DD has been made by \cite{Riesgo2006} providing the 85 percent probability density ellipse. PNe with log(\ha/\nitrogen)$<$0 and log(\ha/\sulfurt)$<$0.4 (below the region occupied by supernova remnants, hereafter SNRs) may experience some shock activity \citep[e.g.][]{Riesgo2006,bohigas2008,akras2016,akrasLIS2016}. BPT diagrams for PNe, SNRs, and \hii~regions based on observations are presented by \cite{Frew2010,sabin2013}. Very recently, the so-called \lq\lq 2D extended diagnostic diagram \rq\rq\ was used by \citet{Barria2018c,Barria2018a,Barria2018b} to search for shocked regions or identify low-ionization structures in PNe.

\subsubsection{Log(\ha/\sulfurt) versus log(\ha/\nitrogen)}

Figure~\ref{DD1} displays the SMB log(\ha/\nitrogen) versus log(\ha/\sulfurt) DD for Abell~14. The line ratios from each individual spaxel (cyan dots), the 1D simulated long-slit spectra [observed: purple circles and modelled \citep{akras2016}: red triangles] and the 1D photoionization models from the 3MdB  \citep[open black squares, ][]{Morisset2015}  are presented. The size of the open black squares corresponds to the logU\footnote{The dimensionless parameter U is also known as ionization parameter and it is defined as the ratio of the hydrogen-ionizing photon number to the total-hydrogen density.} parameter of each {\sc cloudy} model (the bigger the symbol the higher the logU parameter). The inner plot illustrates the distribution of the observed and modelled \citep{akras2016} simulated long-slit emission-line ratios as a function of the slits' position angle from -90 to 90\degree. The black cross indicates the mean value of the corresponding line ratios obtained from all the 1D simulated long-slit VIMOS spectra.

Abell~14 lies in the left-bottom corner and it is characterized by 0.4$<$log(\ha/\sulfurt)$<$0.6. These values barely indicate shocks in Abell~14. It is worth mentioning that the distribution of individual spaxels is quite different compared to the simulated long-slit ratios. Spaxels' values cover a wider range  in log(\ha/\sulfurt) (from 0 to 0.8) and log(\ha/\nitrogen) (from -1.0 to -0.5). This bias to lower ratios from the individual spaxels and higher ratios from the integrated 1D long-slit spectra may lead to the misinterpretation of excitation mechanism: in the first case, towards the existence of shocks and in the second case towards the excitation by the UV stellar radiation field. This delicate point has also been pointed out by \cite{Morisset2018}. 

The comparison between the line ratios derived from the 1D simulated long-slits of the VIMOS datacube and those from the 3D photoionization model \citep{akras2016} developed using the {\sc mocassin}\footnote{MOnte CArlo SimulationS of ionized Nebulae  \citep{Ercolano2003,Ercolano2005}.} code reveals a significant difference of the order of 0.3-0.4~dex. Moreover, one can see that photoionization models can reproduce very low line ratios [log(\ha/\sulfurt)$\sim$-0.4 and log(\ha/\nitrogen)$\sim$-1], without considering shocks (Figure~\ref{DD1}). The common characteristic of this group of models is the very low logU parameter ($<$-3.4), i.e., very low-excitation nebulae. It is also evident that both line ratios, \ha/\nitrogen\ and \ha/\sulfurt, decrease (i.e. the \nitrogen\ and \sulfurt\ lines become stronger) with the decrease of logU (smaller squares). Therefore, the log(\ha/\nitrogen) versus log(\ha/\sulfurt) DD itself is not enough to distinguish nebular gas excited by shocks or UV stellar radiation. The criterion, log(\ha/\sulfurt)$<$0.4,  which commonly applied for supernova remnants \citep[e.g.][]{leonidaki2013}, is not adequate to argue for the presence of shocks in PNe.

\subsubsection{Log(\sulfurt/\ha) versus log(\oxygeniii/\hb)}

Figure~\ref{DD2} presents the classical BPT diagram log(\sulfurt/\ha) versus log(\oxygeniii/\hb). The majority of the {\sc cloudy} models with logU$>$-3.4 lie in the regime where PNe are expected \citep{Frew2010,sabin2013}. All these models have 0.6$<$log(\oxygeniii/\hb)$<$1.5 and -2.5$<$log(\sulfurt/\ha)$<$-0.4. For the very low-excitation models (logU$<$-3.4), log(\oxygeniii/\hb) is lower than 0.5, while log(\sulfurt/\ha) varies between -1.5 and 0.5.

Spaxels, on the other hand, cover a wider range of values, -0.5$<$log(\oxygeniii/\hb)$<$0.70 and -0.75$<$log(\sulfurt/\ha)$<$0.1. According to these values, it could be suggested that some regions in Abell~14 are UV excited, while some others are affected by shock interactions [cyan points with log(\sulfurt/\ha)>-0.4 occupy the same area with young and evolved SNRs \citep{Frew2010,sabin2013}]. However, all these models lie within the PNe limits (brown lines). Photoionization {\sc cloudy} models with logU$<$-3.4 can successfully produce low log(\oxygeniii/\hb) and high log(\sulfurt/\ha) line ratios. The line ratios obtained from the 1D simulated long-slit spectra indicate that Abell~14 lies in a transition region between PNe and SNRs with no extreme values like those obtained from individual spaxels. 

\subsubsection{Log(\nitrogen/\ha) versus log(\oxygeniii/\hb)}
The log(\nitrogen/\ha) versus log(\oxygeniii/\hb) DD of Abell~14 is illustrated in Fig.~\ref{DD3}. Similarly to the previous diagram, log(\nitrogen/\ha) also increases as logU decreases, and its maximum value is close to 0.8. Abell~14 lies at the outskirts of the grid of {\sc cloudy} models consistent with very low-ionization models (logU$<$-3.4). It is evident that Abell~14 is not characterized by line ratios typical to PNe.

\subsubsection{Log(\nitrogena/\hb) versus log(\oxygeniii/\hb)}

Although strong \nitrogena\ and \oxygeni\ lines are usually attributed to shock activity \citep[e.g.][]{phillps1998}, there is no proper DD of these lines for PNe. These lines are not always detected in PNe (or they are very weak), and this makes very difficult to construct a trustworthy DD from observations. A grid of photoionization models, as the Mexican Million Models, may be of great help in such an analysis. 

Figure~\ref{DD4} displays the log(\nitrogena/\hb) versus log(\oxygeniii/\hb) DD. Both the individual spaxels and the 1D simulated long-slit spectra show an overlap with {\sc cloudy} models only for logU between -4.4 and -3.4, and \nitrogena/\hb\ between -0.75 and -0.25. Note that there are spaxels with \nitrogena/\hb$>$1, which could be easily interpreted as shock collisions, even if they can be reproduced by photoionization models with very low logU.

The observed and modelled line ratios display a difference in log(\nitrogena/\ha) of almost 1.5~dex (see the inner plot in Figure~\ref{DD4}). Recall that the {\sc mocassin} model of Abell~14 does not reproduce the intensity of \nitrogena~$\lambda$5200. It can also be observed that the eastern part (PA$<$0, blue points) has systematically higher log(\nitrogena/\hb) compared to the western part (PA$>$0, red points) of $\sim$0.3~dex. This may be associated with the orientation of the nebula, resulting in a fainter \nitrogena~$\lambda$5200 line in the western part (see Fig.~\ref{linemaps}).

\subsection{The 3MdB grid}

In order to verify if there is any additional thermal mechanism as suggested by \citet{akras2016}, we searched the 3MdB for the {\sc cloudy} model(s) that reproduces simultaneously the log(\oxygeniii/\hb), log(\sulfurt/\ha), and log(\nitrogen/\ha) line ratios, considering a variance equal to the errors of their observed values (see section 4.5). Two models, a radiation bounded and an 80\% matter-bounded model, with negligible differences in line ratios and stellar parameters, are found. For our analysis below, we use the 80\% matter-bounded model. 

For this best model, we get $T_{\rm{eff}}$=150\,000~K, log(L/L$_{\odot}$)=3.0, and logU=-3.8, i.e., a hot and faint evolved white dwarf. The size of the nebula (r$_{\rm{out}}$) is $\sim$10$^{17}$~cm with a low electron density ($<$100~cm$^3$). All these parameters are consistent with the results from our previous study of Abell~14 \citep[][]{akras2016}. A very good agreement is also found for the remaining emission-line ratios: log(\oxygeniii/\hb) (modelled=0.46; observed=0.42), log(\sulfurt/\ha) (-0.43; -0.40), log(\nitrogen/\ha) (0.82; 0.79), log(\nitrogen~5755/\ha) (-0.62; -0.62), log(\heliumb/\hb) (-0.61; -0.58), and log(\helium/\hb) (-0.77; -0.72). Therefore, most of the emission-line ratios of Abell~14 can be reproduced from the UV stellar radiation field.

However, a notable difference is found in the log(\nitrogena/\hb) (-1.10; -0.39) and log(\oxygenii/\hb) (0.25; 0.60) line ratios. The deviation of these two lines from the pure photoionization models cannot be explained without considering an additional thermal mechanism. Log(\oxygeni/\ha) ratio is also found to be inconsistent with the 3MdB models (0.04; -1.27\footnote{The \oxygeni~$\lambda$6300 line is very weak and no line map is constructed. This value has been taken from the integrated spectrum after using the VIMOS datacube as direct input to the {\sc alfa} code.}). Such a bright \oxygeni~$\lambda$6300 line should have easily been observed. On the contrary, our VIMOS data indicate a much weaker emission.

Two possible solutions were explored to explain the high \nitrogena/\hb, \oxygenii/\hb, and \oxygeni/\ha\ line ratios: i) the effect of the stopping criteria on the emission-line ratios in the {\sc cloudy} simulations \citep[see][]{bohigas2008}, but it does not explain the observed discrepancy, and ii) the high O abundance of the 3MdB {\sc cloudy} models, $\sim$1~dex more abundant than the observations indicate. Such difference in O abundance, the most important coolant element, will significantly alter the ionization structure of the model and the rest of the emission-lines ratios.

\begin{table*}
\caption[]{~Nebular parameters of Abell~14 obtained from the integrated spectrum and the 1D simulated long-slits spectra for PA=-90,-80,-20, and 0$^{\circ}$} \label{table3}
\begin{tabular}{cccccccccc}
\hline
Ion & $I \left( \lambda \right)^{\dag}$ & $I \left( \lambda \right)^{\dag\dag}$ & \multicolumn{4}{c}{PA} & B03$^a$  & H10$^b$  \\

 & {\sc neat} & Maps  & -90$^{\circ}$ & -80$^{\circ}$ & -20$^{\circ}$ & 0$^{\circ}$ &  &    \\
 \hline                                                                          
  He$^+$/H  & 1.3 $\pm$ 0.3(-1)  & 1.4(-1)   & 0.8(-1)  &  0.9(-1) & 1.3(-1) & 1.3(-1) & 1.19(-1) & 1.72(-1) \\
  He$^{2+}$/H & 2.0 $\pm$ 0.2(-2)  & 2.1(-2) & 3.7(-2)  &  3.6(-2) & 2.1(-2) & 2.0(-2) & 3.70(-2) &  2.40(-2)\\
  He/H  & 1.5 $\pm$ 0.3(-1)  & 1.6(-1)   & 1.2(-1)  &  1.3(-1) & 1.5(-1) & 1.5(-1) & 1.56(-1) & 1.96(-1)\\
  N$^+$/H  & 1.83 $\pm$ 0.25(-4)  & 1.55(-4)   & 1.01(-4)  &  1.11(-4) & 2.25(-4) & 1.99(-4) &  1.80(-4) & 2.16(-4)\\
  ICF(N) & 1.77 $\pm$ 0.13  & 1.80  & 2.66   &  2.46  & 1.96  & 1.55  & -    & 4.84\\
  N/H  & 3.26 $\pm$ 0.48(-4)  & 2.79(-4)     & 2.66(-4)  &  2.71(-4) & 4.41(-4) & 3.08(-4) & 3.90(-4) &  1.050(-3)\\
  O$^+$/H  & 11.9 $\pm$ 3.4(-5)  & 8.9(-5)    & 6.1(-5)  &  7.6(-5) & 1.3(-4) & 1.7(-4) & -         & 5.35(-5) \\
  O$^{2+}$/H & 7.94 $\pm$ 1.31(-5)  & 6.19(-5)  & 8.28(-5)  &  9.13(-5) & 11.10(-5) & 7.56(-5) & 4.8(-5) & 16.90(-5) \\
  ICF(O) & 1.09 $\pm$ 0.02  & 1.08  & 1.26   &  1.22  & 1.09  & 1.09 & -         & 1.14\\
  O/H  & 2.17 $\pm$ 0.48(-4)  & 1.64(-4)     & 1.81(-4)  &  2.04(-4) & 2.64(-4) & 2.68(-4) & 1.30(-4) & 2.54(-4)\\
  S$^+$/H  & 2.18 $\pm$ 0.30(-6)  & 1.91(-6)    & 1.23(-6)  &  1.23(-6) & 2.64(-6) & 2.22(-6) & 1.33(-6) & 1.77(-6) \\
  ICF(S) & 3.14 $\pm$ 0.25  & 3.21  & 4.19   &  3.95  & 3.49  & 2.73 & -        & 1.19 \\
  S/H   & 6.9 $\pm$ 1.1(6)  & 6.1(-6)    & 5.2(-6)  &  5.0(-6) & 9.2(-6) & 6.1(-6) & 9.3(-6) & 3.15(-6)\\
  Ar$^{2+}$/H & 1.58 $\pm$ 0.25(-6)  & 1.29(-6) & 1.30(-6)  &  1.20(-6) & 1.85(-6) & 1.46(-6) & 1.14(-6) & 1.02(-6)\\
  ICF(Ar) & 1.19 $\pm$ 0.04  & 1.18 & 1.34   &  1.28  & 1.16  & 1.27  & -         & 1.44\\
  Ar/H  & 1.88 $\pm$ 0.31(-6)  & 1.53(-6)    & 1.73(-6)  &  1.53(-6) & 2.14(-6) & 1.85(-6) & 2.10(-6) & 1.46(-6)\\ 
  Ne$^{2+}$/H & 7.28 $\pm$ 2.15(-5)  & 5.18(-5)  & 5.80(-5) & 6.24(-5)& 8.43(-5) & 5.09(-5) & $-$  & 6.55(-5) \\
  ICF(Ne) & 1.09 $\pm$ 0.02  & 1.09  & 1.26 & 1.22  & 1.09  & 1.09 & $-$ &  1.49\\ Ne/H  & 7.96 $\pm$ 2.35(-5)  & 5.63(-5)  & 7.31(-5) & 7.63(-5) & 9.20(-5) & 5.55(-5) & $-$ & 9.79(-5)\\
\hline
  $T_{\rm{e}}$(K) & 11000 $\pm$ 500 &  11532  & 12194 & 11863  &  9912    & 10589  & 12000  & 13270 \\ 
  $n_{\rm{e}}$(cm$^3$) & 10 $\pm$ 40  &  32  & 118   &    10   &    82     &   21  & 85  &   55\\
\hline
\end{tabular}
\begin{flushleft}
$^\dag$ Nebular parameters calculated from the observed portion of Abell~14 observed using the code {\sc neat} (integrated spectra).\\
$^{\dag\dag}$ Nebular parameters calculated from the observed portion of Abell~14 observed using the line maps (integrated spectra).\\
$^a$ \citep{Bohigas2003}: Slit width 3~arcsec oriented at the north--south direction (PA=0), $^b$ \citep{Henry2010}: Slit width 2~arcsec oriented at PA=100. 
\end{flushleft}
\end{table*}

\subsection{Nebular electron density and temperature}

\begin{figure*}
\includegraphics[scale=0.305]{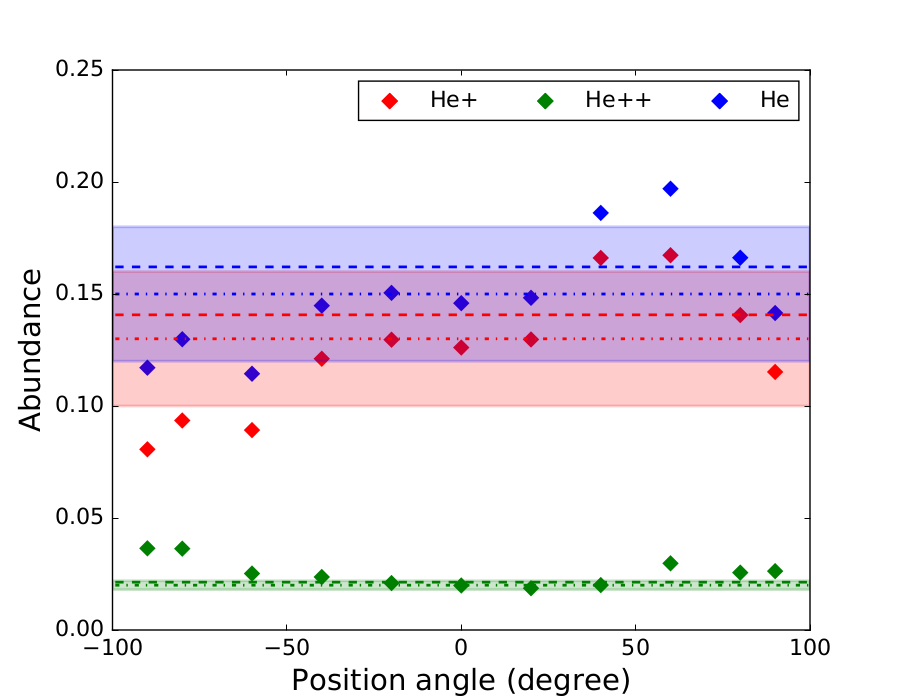}
\includegraphics[scale=0.305]{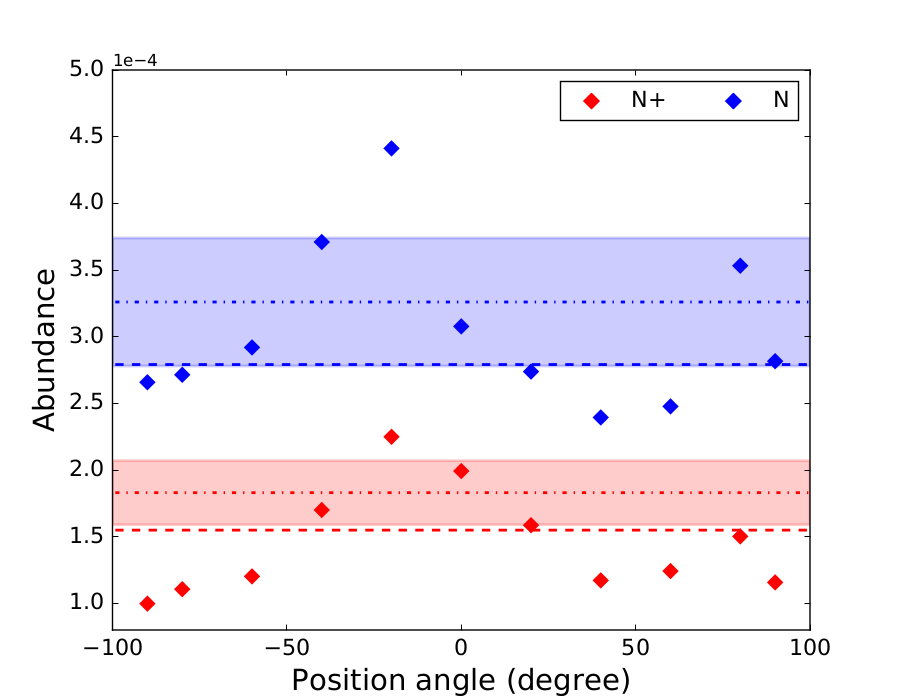}
\includegraphics[scale=0.305]{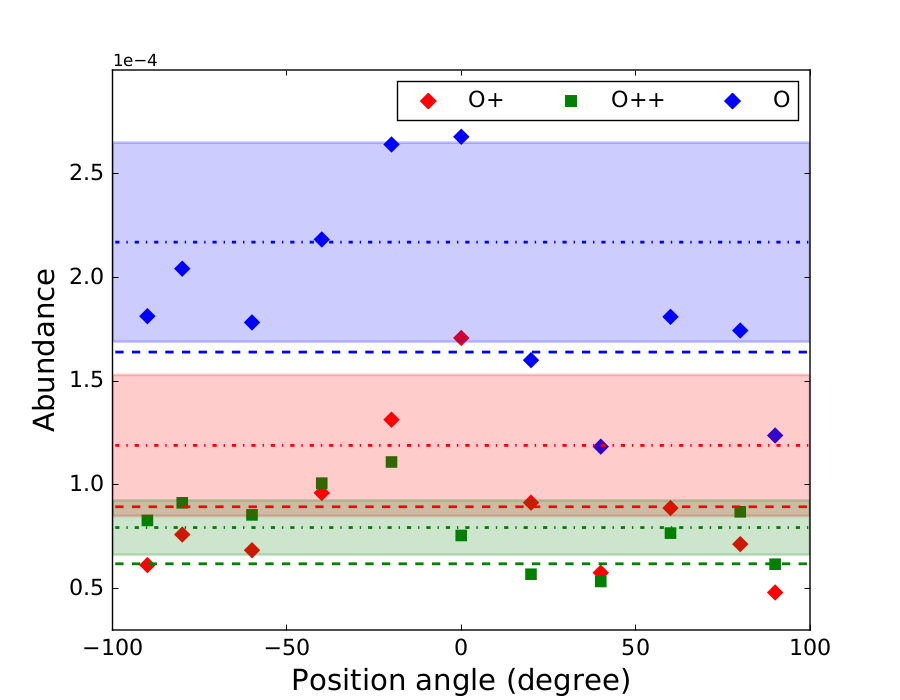}
\includegraphics[scale=0.305]{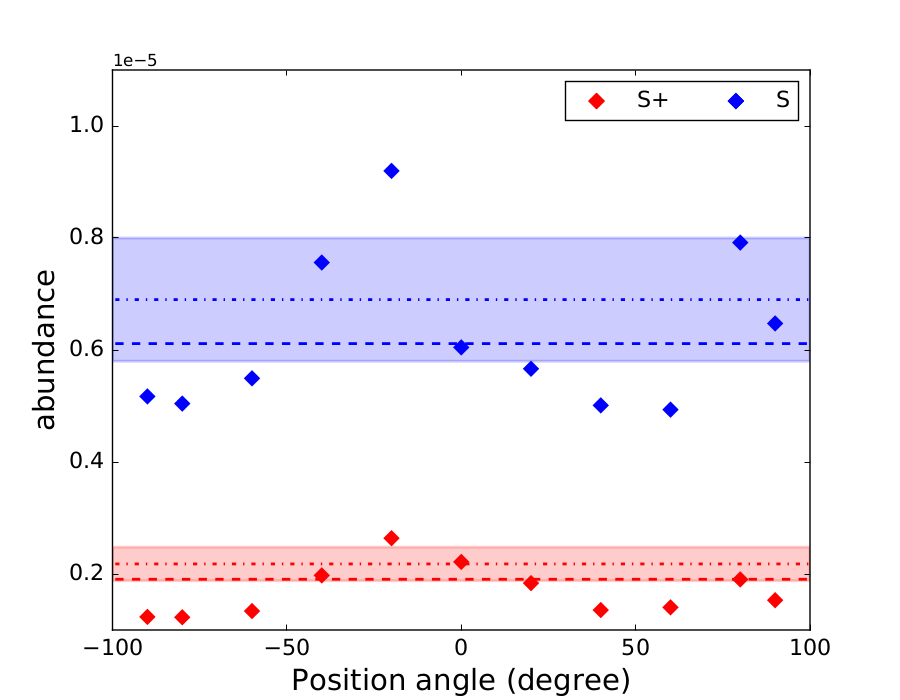}
\includegraphics[scale=0.305]{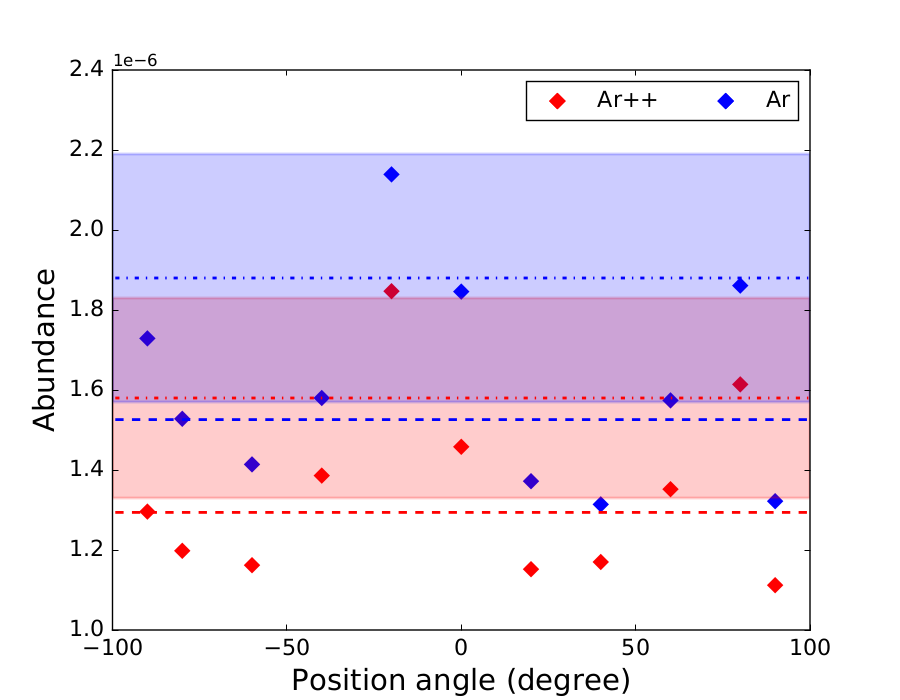}
\includegraphics[scale=0.305]{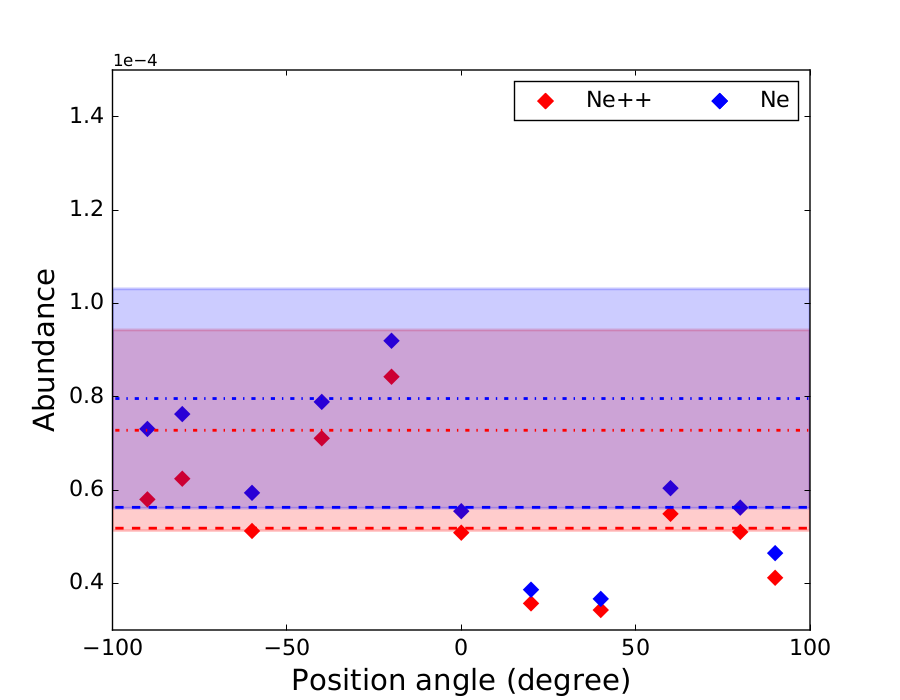}
\caption[]{Ionic and total elemental abundances in the nebula. The points correspond to the simulated spectra from -90 to 90\degree. The dot-dashed and dashed lines are the abundances derived from the integrated spectra using the {\sc neat} code and the line maps, respectively. The coloured areas illustrate the upper and lower bounds of the abundances based on the errors provided by {\sc neat}. The error of each data point is larger than the corresponding error derived for the integrated spectrum.} 
\label{abundances}
\end{figure*}

The electron density and temperature of the nebular gas are also derived using the classical indicators (e.g. $n_{\rm e}$: \sulfurt~6716/6731, $T_{\rm e}$: \nitrogen~6548+6584/5755) available for Abell~14. Both the parameters are derived using the integrated spectrum and the 1D simulated long-slit spectra. The latter allow us to explore for possible variations within the nebula based on the PA of each slit. Table~\ref{table3} lists the resulting values of these two parameters.

\begin{figure*}
\includegraphics[scale=0.305]{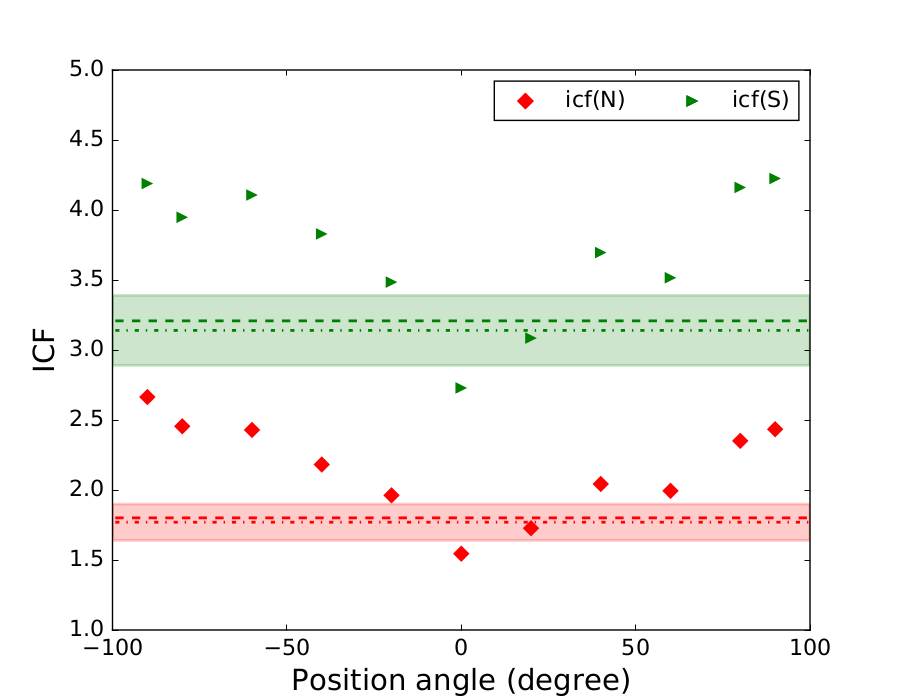}
\includegraphics[scale=0.305]{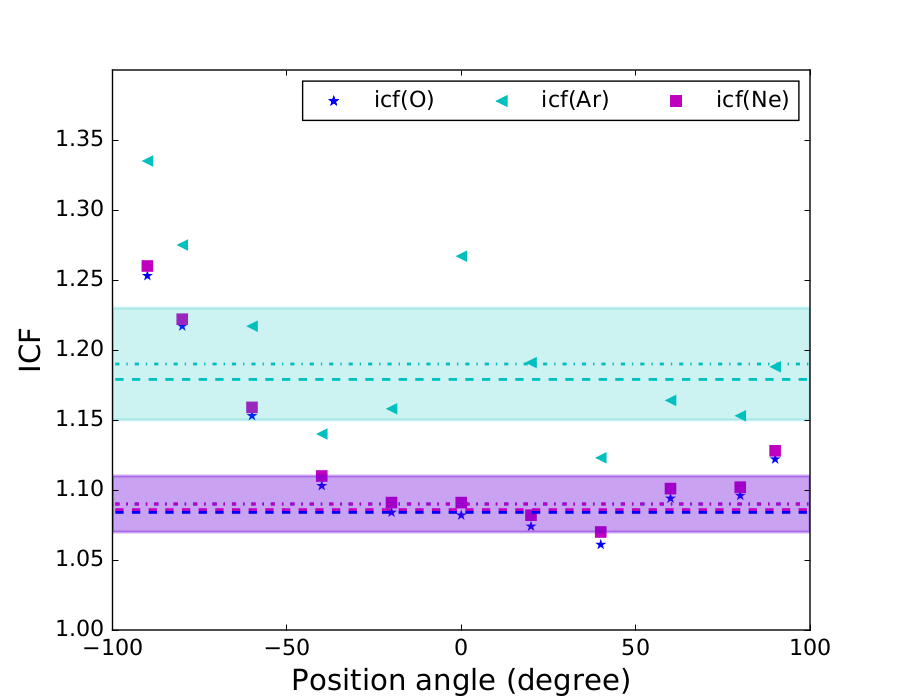}
\includegraphics[scale=0.305]{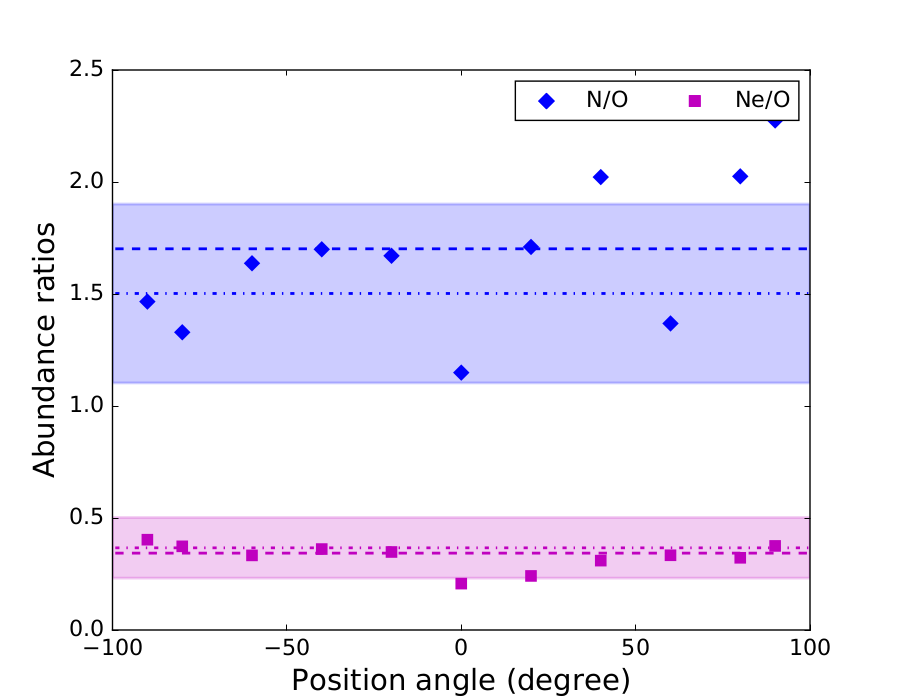}
\includegraphics[scale=0.305]{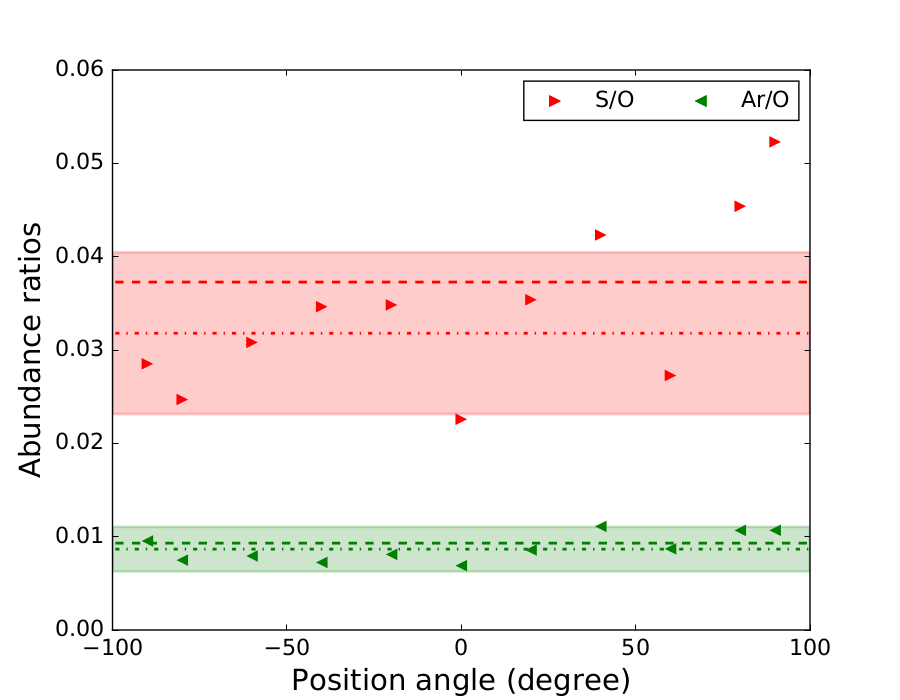}
\caption[]{ICFs and abundance ratios for Abell~14 as a function of the PA of the simulated slits. The symbols and the lines are the same as in Figure~\ref{abundances}.}
\label{icfs}
\end{figure*}

Abell~14 does not show any significant variation in the electron density or  temperature as a function of the PA of slit. Electron temperature varies from 10000 to 12500~K, which is common in PNe, as the recent spatially resolved analysis of NGC~7009 \citep{Walsh2016}, NGC~40 \citep{Ferreira2011} and NGC~2438 \citep{ottl2014} has also shown. The error of the electron temperature derived from the simulated slits is of the order of 2000~K. The integrated spectra obtained from the {\sc alfa} code and the line maps give very similar $T_{\rm e}$ values of 11000 and 11500~K, respectively. Electron density, on the other hand, is found to be systematically lower than 200~cm$^{-3}$, consistent with the previous studies.

\subsection{Ionic and total abundances}

With the $n_{\rm e}$ and $T_{\rm e}$ of the nebular gas calculated, the ionic and total chemical abundances of Abell~14 can also be derived. Given that only a limited number of ions of a given species are detected in the optical wavelengths, the ionization correction factors (ICFs) from \cite{Inglada2014} are used to calculate the total abundances. ICFs are used in order to take into account the contribution of the unobserved ions in the total chemical abundances. The ionic and total abundances of He, N, O, S, Ne and Ar, as well as their corresponding ICFs are given in Table~\ref{table3} for the integrated spectra and 1D four simulated long-slit spectra with the same orientation as the observed spectra in \cite{Bohigas2003} and \cite{Henry2010}. 

Figure~\ref{abundances} displays the ionic and total abundances as a function of the PA of the simulated slits. All chemical abundances, from the integrated {\sc neat} spectrum  (dot-dashed lines), the line maps (dashed lines), and the 1D simulated long-slit spectra (coloured diamonds), were calculated using the code {\sc neat}\footnote{Nebular Empirical Analysis Tool \citep[{\sc neat}; ][]{Wesson2012}.}  The coloured areas indicate the abundance errors for the integrated {\sc neat} spectrum. No errors are given for the abundances derived from the line maps or 1D simulated long-slit spectra but they are always larger than those derived from the integrated {\sc neat} spectrum. Both ionic and total abundances show some variation within the nebula but it is smaller compared to their uncertainties. Moreover, some differences in the chemical abundances derived from the integrated {\sc neat} spectrum and the line maps have been found, which we attributed to the criteria used to determine the line fluxes from the line maps (see Sect.~3).

Variations in the chemical composition of PNe have been reported for a few PNe but their large errors have prevented their confirmation \citep{Ferreira2011,Hektor2013}. The MUSE data of NGC~7009 have also shown spatial variations in the ionic abundances of various elements from 0.5 up to even 2~dex \citep[see][]{Walsh2018}.

In Figure~\ref{icfs}, we present the ICFs and abundances ratios as a function of the PA for the 1D simulated long-slits VIMOS spectra (coloured diamonds), the integrated {\sc neat} spectrum (dot-dashed) and the line maps (dashed lines). For the ICFs, we get a significant variability within the nebula. All the elements except for Ar have higher ICFs in the polar direction and lower in the equatorial region (Figure~\ref{icfs}). 3D photoionization models have demonstrated that the ICFs vary within the nebula \citep[][]{Morisset2017}. Moreover, it has also been suggested that a correction on the ICFs may be needed when we deal with non-spherical nebulae \citep[][]{Goncalves2012} and Abell~14 is one of these cases.

Regarding the N/O, S/O, Ar/O and Ne/O abundance ratios, we find a reasonable agreement between the integrated spectra and the 1D simulated long-sit spectra (Figure~\ref{icfs}). These results denote that the variation found in ICFs is likely real and responsible for the homogeneous abundance ratios. Deep, high quality IFU data from more PNe are needed in order to verify the variation of ICFs and homogeneity in the abundance ratios.

Based on the chemical abundances, Abell~14 is classified as a Type~I PN \citep{Peimbert1978}. The high log(N/O) ratio ({\sc alfa}, 0.18; maps, 0.23) and He abundance (0.15 and 0.16) are comparable with those found in bipolar PNe \citep{Pottasch2010} and they are consistent with the prediction of high metallicity evolutionary models of massive PN progenitors \citep{Karakas2010}. 

\subsection{Comparison with previous long-slit spectra}
Our methodology allows us to directly compare our results with those obtained from previous long-slit spectroscopy. The 1D simulated long-slit spectrum at PA=0\degree\ can be directly compared with the observed spectrum from \cite{Bohigas2003}. A good agreement is found for the \hb\ flux, as well as for the total abundances of He, N, and Ar, except O and S. Our O abundance is almost twice Bohigas's value, while S is almost 35~percent less abundant. \cite{Bohigas2003} does not provide the abundance for the singly ionized O, because \oxygenii\ $\lambda$3727 is not detected in his spectrum. Our O$^{+}$ abundance is almost two times larger than O$^{++}$ (Table~\ref{table3}), and it can explain the discrepancy in the O total abundance.

The observed spectrum from \cite{Henry2010} can be compared with the simulated slit at PA=-80\degree. The two spectra are found to be very different. First of all, the \hb\ flux reported by \cite{Henry2010} does not agree with the fluxes of any of our simulated spectra. A significant difference is also observed between the fluxes of the doublet lines \nitrogen\ and \sulfurt, which are 30-50~percent weaker in our data. He is also found $\sim$33 percent less abundance compared to the previous studies.

The very high N abundance reported by \cite{Henry2010} is not supported by our VIMOS data. The different formulae of ICFs used in this work as well as the different oxygen ionic abundances may be responsible for this discrepancy in N abundances. For instance, \cite{Henry2010} estimated ICF(N)=4.8 using the expressions of \cite{Kingsburgh1994} while our estimations using the formula of \cite{Inglada2014} are between 1.5 and 2.5. The high ratio of O/O$^{+}$=5.1 \citep{Henry2010} results in the high ICF(N). 

The total O and S abundances are also found to be significantly different. Our O abundance is 20~percent lower than the value derived by \citet{Henry2010}, while S is 40~percent more abundant. This discrepancy in the abundance of S is also attributed to the different O ionic abundances. The high O$^{2+}$/O$^+$ (3.2) and low O$^+$/O (0.2) ratios reported by \cite{Henry2010} yield a low ICF(S), and consequently a low total abundance for S. Our O$^{2+}$/O$^+$ and O$^+$/O ratios derived from the PA=-80\degree\ simulated long-slit spectrum are determined 1.20 and 0.37, respectively. Even if the ICFs from \cite{Kingsburgh1994} were used, a higher ICF(S) and total S abundance should be obtained considering our ionic abundances.

\subsection{The apparent central star}

\begin{figure*}
\includegraphics[scale=0.670]{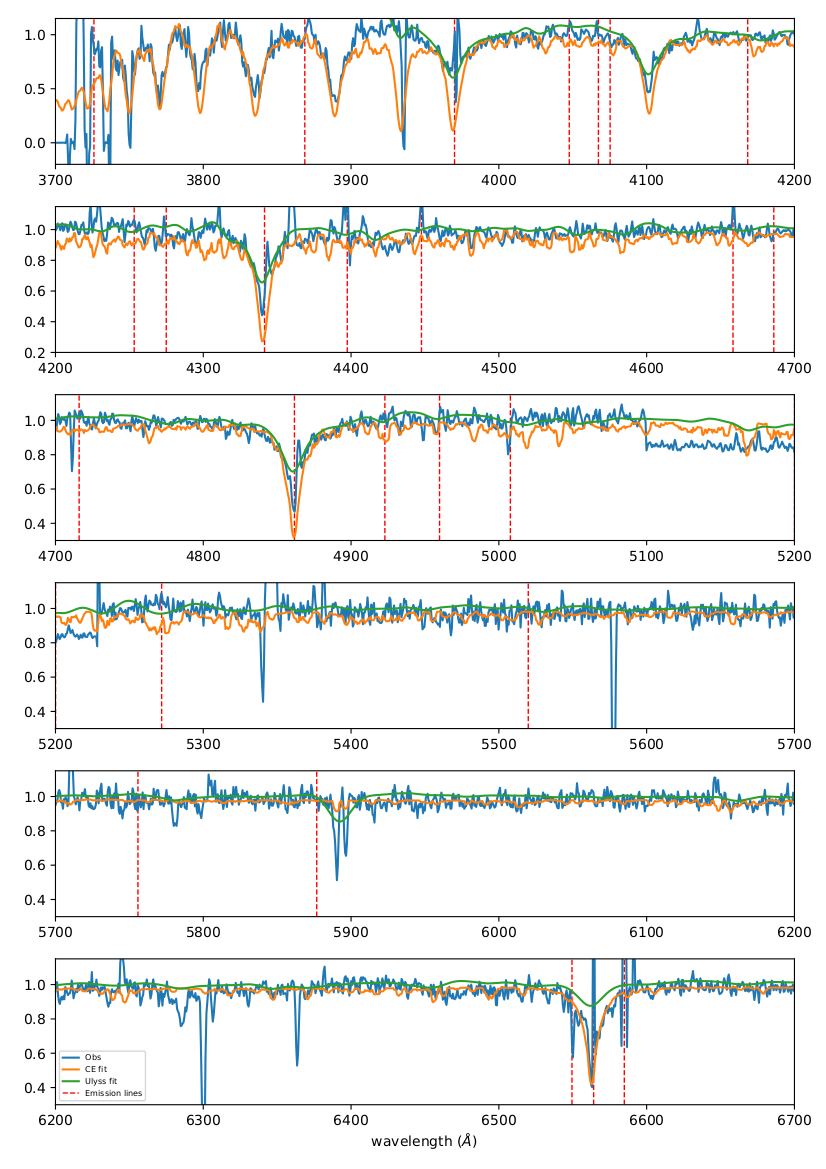}
\caption[]{Central star spectrum and fitted models. In blue we see the observed spectrum, in orange the CE fit and in green the Ulyss fit. The red dashed lines mark the positions of the measured emission-lines of the nebula. emission-lines are not removed.}
\label{CS_fit}
\end{figure*}

One of the advantages of observing PNe with IFS is that besides the nebula, the central star(s) (if detected) can also be studied. In the case of Abell~14, we studied its apparent central star, i.e., the star at the geometrical centre of the nebula. From our datacube, we extracted its spectrum using a 2D Gaussian model fitted along the dispersion axis. The stellar integrated spectrum is presented in Figure~\ref{CS_fit}. Our results indicate that this star cannot be the ionizing star of Abell~14, as we will discuss in the following.

Two different methods were used to obtain the spectral classification of the central star. In the first, we compared the spectrum of the star with the grid of theoretical spectra from \citet{Coelho2014}. This theoretical library consists of 3727 spectra with effective temperature ranging from 3000 to 25000~K, log(g) from -0.5 to 5.5 and metallicity Fe/H from -1.3 to +0.2. We wrote an IDL code to obtain an interpolated spectrum from the grid and fit it to the observed VIMOS data using the global optimization method called Cross Entropy \citep[(CE); further details of the method can be found in][]{Monteiro2010}. With this method, we estimated values for the stellar atmospheric parameters obtaining $T_{\rm{eff}} = $7685$\pm$2937~K, log(g) = 1.7$\pm$1.0~$cm/s^{2}$ and log(Fe/H) = 0.3$\pm$0.3.

We validated the results using a second well-established procedure. We used the code {\sc ulyss} (Universit\'{e} de Lyon Spectroscopic analysis Software) developed by \cite{Koleva2009}. The code obtains the atmospheric parameters by fitting the observed spectrum with an interpolated spectrum of the empirical library ELODIE \citep{Prugniel2001,Prugniel2007}. The fitting procedure uses $\chi^{2}$ minimization.

The fits obtained with {\sc ulyss} resulted in the atmospheric parameters $T_{\rm{eff}} =$ 7909$\pm$135~K, log(g) = 1.4$\pm$0.1~$cm/s^{2}$ and log(Fe/H) = -1.7$\pm$0.1. While the effective temperature and surface gravity are in good agreement with the value we obtained with the first method, the metallicity is significantly different.

In Figure~\ref{CS_fit}, we show the observed spectrum and both fitted spectra discussed above. The {\sc ulyss} fit does not reproduce the major absorption lines as well as our code. Our code also manages to fit further to the blue. The star has been previously classified by \citet{Weidmann2011} as a B8-9 type due to the presence of Balmer and He absorption lines. Our analysis indicates an A5 type star. In any case, such a star is not able to ionise a nebula as Abell~14. We note also that the stellar absorption lines are aligned with the nebula emission-lines (i.e., both show the similar radial velocities), as shown in Fig.~\ref{CS_fit}.

The probability that the central star of Abell~14 is aligned with a field star at such small distance is extremely low. If this alignment is true, we must agree with Henry Boffin that \lq\lq {\it Nature likes to play tricks with us!} \rq\rq \citep{Boffin2018}.

The star has also been observed by the {\it Gaia} satellite. Its parallax is provided in the second data release ($\pi$=0.1948$\pm$0.036, \citealt{gaia2018}). The low fractional error of the parallax measure ($\sigma_\pi$/$\pi$~=~0.18) allows us to get the distance from the inverse of parallax, D~=~1/$\pi$~=~5.1$\pm$0.9~kpc.  
For comparison, the distance of this star has also been derived using a robust probabilistic approach \cite{bailer2018} and it is found to be 4.34~kpc with 1$\sigma$ minimum and maximum values of 3.74 and 5.15~kpc, respectively. Note that \citet{akras2016} also got a distance of 4~kpc from their best-fitting 3D model.

It is  worth mentioning here that a systematic offset in the {\it Gaia} DR2 parallaxes has been reported by several authors. This offset is likely between 10 and 100~mas, strongly depending on the magnitudes and colours of the sources as well as their position on the sky \citep{Lindegren2018}. More specifically, the offset has been estimated 0.029~mas from a sample of quasars \citep{Lindegren2018}, 0.046 and 0.048~mas from a sample of Cepheids \citep{Riess2018,Groenewegen2018}, 0.057~mas from a sample of RR-Lyrae stars \citep{Muraveva2018}, 0.082~mas from eclipsing binaries. A year later, more studied concluded to a non-negligible parallax offset in the {\it Gaia} DR2 [0.031~mas, \cite{Graczyk2019}; 0.054~mas, \cite{Schonrich2019}; 0.0523~mas, \cite{Leung2019}; 0.0517~mas, \cite{Khan2019}; 0.0528~mas, \cite{Zinn2019}; 0.03838~mas, \cite{Hall2019}; 0.075~mas, \cite{Xu2019}].

It is not evident which parallax offset is applicable for the central star of Abell~14. Therefore, the lowest (0.029~mas), the highest (0.082~mas) and the weighted average from all the aforementioned parallax offsets (0.051~mas) were considered to correct its parallax and distance \citep[see also][]{Gordillo2020}. These offsets yield distances of 4.46, 3.6, and 4.08~kpc, respectively, and they agree with the 4~kpc distance derived from 3D modelling \citep{akras2016}. We argue that the A-type star at the centre of Abell~14 is likely a companion star in a binary system.

\section{Discussion and Conclusions}

The ionization structure and chemical composition of Abell~14 were studied using new the VLT@VIMOS IFU data. This new data set allowed us to investigate how emission-line ratios are distributed in two spatial directions following three approaches: (1) the integrated spectrum, (2) spaxel-by-spaxel spectra, and (3) 1D simulated long-slit spectra in various position angles.

The BPT and SMB diagnostic diagrams were constructed using our VIMOS data and photoionization models. A spaxel-by-spaxel analysis of the emission-lines in the BPT diagrams may lead to misinterpretation of the results, since they cover a wider range of values and substantially different from those derived by the 1D simulated long-slit spectra. The common log(\sulfurt/\ha)$<$0.4 criterion to trace shock-excited gas is not directly applicable for PNe. The contribution of the UV radiation from the central star and shock interactions cannot be distinguished using only emission-line ratios.

The 3MdB grid of photoionization models was also explored in order to seek for those models that are capable to reproduce high line ratios usually attributed to shock excitation. Despite the photoionization models have a spherically symmetry, we found two models (one radiation bounded and one 80~percent matter bounded), which can reproduce, within the uncertainties, the observed line ratios. These models indicate a very large and evolved nebula, with a low -luminosity central star in agreement the previous studies. However, the log(\nitrogena/\hb) and log(\oxygenii/\hb) line ratios derived from the observations and the models were significant different, and an additional thermal mechanism may be responsible for this discrepancy.

The ionic and total chemical abundances obtained from the 1D simulated long-slit spectra indicate some variation with the slit's position angle. However, the large abundance errors did not allow us to confirm this behaviour for Abell~14. On the other hand, the abundance ratios (N/O, S/O, Ar/O and Ne/O) did not show any variation with the slit's position angles. That result was attributed to the variability of ICFs, which were found to increase from the equator to the polar direction.

Besides the analysis of the nebular gas in distinct spatial directions, IFS also allowed us to study the spectral properties of the star at the geometric centre of the nebula. The star was classified as an A-type star with $T_{\rm eff}$~=~7909$\pm$135~K and log($g$)~=~1.4$\pm$0.1~cm~s$^{-2}$, and it could not be the ionizing source of the nebula. This star and the nebula were found to located approximately at the same distance of 4~kpc, and we concluded that the central star of Abell~14 is very likely a binary system.

\section*{Acknowledgements}
The authors would like to thank the anonymous referee for the valuable comments and constructive suggestions. SA acknowledges the financial support from the Brazilian agencies CNPq (grant 454794/2015-0) and CAPES for a fellowship from the National Postdoctoral Program (PNPD). IA acknowledges the support of CAPES, Ministry of Education, Brazil, through a PNPD fellowship. The authors wish to thank the VIMOS support staff at Paranal for taking these service mode observations [program ID: 098.D-0436(A)]. This research was performed using the facilities of the Laborat\'orio de Astrof\'isica Computacional da Universidade Federal de Itajub\'a (LAC-UNIFEI). Finally, this publication makes use of many software packages in Python, including: Matplotlib \citep{Hunter2007}, NumPy \citep{Walt2011}, SciPy \citep{SciPy2020} and AstroPy Python \citep{Astropy2013,Muna2016,Astropy2018}.





\bibliographystyle{mnras}  
\bibliography{references}

\begin{thebibliography}{}
\makeatletter
\relax
\def\mn@urlcharsother{\let\do\@makeother \do\$\do\&\do\#\do\^\do\_\do\%\do\~}
\def\mn@doi{\begingroup\mn@urlcharsother \@ifnextchar [ {\mn@doi@}
  {\mn@doi@[]}}
\def\mn@doi@[#1]#2{\def\@tempa{#1}\ifx\@tempa\@empty \href
  {http://dx.doi.org/#2} {doi:#2}\else \href {http://dx.doi.org/#2} {#1}\fi
  \endgroup}
\def\mn@eprint#1#2{\mn@eprint@#1:#2::\@nil}
\def\mn@eprint@arXiv#1{\href {http://arxiv.org/abs/#1} {{\tt arXiv:#1}}}
\def\mn@eprint@dblp#1{\href {http://dblp.uni-trier.de/rec/bibtex/#1.xml}
  {dblp:#1}}
\def\mn@eprint@#1:#2:#3:#4\@nil{\def\@tempa {#1}\def\@tempb {#2}\def\@tempc
  {#3}\ifx \@tempc \@empty \let \@tempc \@tempb \let \@tempb \@tempa \fi \ifx
  \@tempb \@empty \def\@tempb {arXiv}\fi \@ifundefined
  {mn@eprint@\@tempb}{\@tempb:\@tempc}{\expandafter \expandafter \csname
  mn@eprint@\@tempb\endcsname \expandafter{\@tempc}}}

\bibitem[\protect\citeauthoryear{{Akashi}, {Bear}  \& {Soker}}{{Akashi}
  et~al.}{2018}]{akashi2018}
{Akashi} M.,  {Bear} E.,   {Soker} N.,  2018, \mn@doi [\mnras]
  {10.1093/mnras/sty029}, \href
  {http://cdsads.u-strasbg.fr/abs/2018MNRAS.475.4794A} {475, 4794}

\bibitem[\protect\citeauthoryear{{Akras} \& {Gon{\c c}alves}}{{Akras} \&
  {Gon{\c c}alves}}{2016}]{akrasLIS2016}
{Akras} S.,  {Gon{\c c}alves} D.~R.,  2016, \mn@doi [\mnras]
  {10.1093/mnras/stv2139}, \href
  {http://adsabs.harvard.edu/abs/2016MNRAS.455..930A} {455, 930}

\bibitem[\protect\citeauthoryear{{Akras}, {Boumis}, {Meaburn}, {Alikakos},
  {L{\'o}pez}  \& {Gon{\c c}alves}}{{Akras} et~al.}{2015}]{Akras2015}
{Akras} S.,  {Boumis} P.,  {Meaburn} J.,  {Alikakos} J.,  {L{\'o}pez} J.~A.,
  {Gon{\c c}alves} D.~R.,  2015, \mn@doi [\mnras] {10.1093/mnras/stv1468},
  \href {http://cdsads.u-strasbg.fr/abs/2015MNRAS.452.2911A} {452, 2911}

\bibitem[\protect\citeauthoryear{{Akras}, {Clyne}, {Boumis}, {Monteiro},
  {Gon{\c c}alves}, {Redman}  \& {Williams}}{{Akras} et~al.}{2016}]{akras2016}
{Akras} S.,  {Clyne} N.,  {Boumis} P.,  {Monteiro} H.,  {Gon{\c c}alves} D.~R.,
   {Redman} M.~P.,   {Williams} S.,  2016, \mn@doi [\mnras]
  {10.1093/mnras/stw038}, \href
  {http://adsabs.harvard.edu/abs/2016MNRAS.457.3409A} {457, 3409}

\bibitem[\protect\citeauthoryear{{Ali} \& {Dopita}}{{Ali} \&
  {Dopita}}{2017}]{Ali2017}
{Ali} A.,  {Dopita} M.~A.,  2017, \mn@doi [\pasa] {10.1017/pasa.2017.30}, \href
  {http://cdsads.u-strasbg.fr/abs/2017PASA...34...36A} {34, e036}

\bibitem[\protect\citeauthoryear{{Ali} \& {Dopita}}{{Ali} \&
  {Dopita}}{2019}]{Ali2019}
{Ali} A.,  {Dopita} M.~A.,  2019, \mn@doi [\mnras] {10.1093/mnras/stz201},
  \href {http://cdsads.u-strasbg.fr/abs/2019MNRAS.484.3251A} {484, 3251}

\bibitem[\protect\citeauthoryear{{Ali}, {Amer}, {Dopita}, {Vogt}  \&
  {Basurah}}{{Ali} et~al.}{2015}]{Ali2015}
{Ali} A.,  {Amer} M.~A.,  {Dopita} M.~A.,  {Vogt} F.~P.~A.,   {Basurah} H.~M.,
  2015, \mn@doi [\aap] {10.1051/0004-6361/201526223}, \href
  {http://cdsads.u-strasbg.fr/abs/2015A%26A...583A..83A} {583, A83}

\bibitem[\protect\citeauthoryear{{Ali}, {Dopita}, {Basurah}, {Amer}, {Alsulami}
   \& {Alruhaili}}{{Ali} et~al.}{2016}]{Ali2016}
{Ali} A.,  {Dopita} M.~A.,  {Basurah} H.~M.,  {Amer} M.~A.,  {Alsulami} R.,
  {Alruhaili} A.,  2016, \mn@doi [\mnras] {10.1093/mnras/stw1744}, \href
  {http://cdsads.u-strasbg.fr/abs/2016MNRAS.462.1393A} {462, 1393}

\bibitem[\protect\citeauthoryear{{Astropy Collaboration} et~al.,}{{Astropy
  Collaboration} et~al.}{2013}]{Astropy2013}
{Astropy Collaboration} et~al., 2013, \mn@doi [\aap]
  {10.1051/0004-6361/201322068}, \href
  {http://cdsads.u-strasbg.fr/abs/2013A%26A...558A..33A} {558, A33}

\bibitem[\protect\citeauthoryear{{Astropy Collaboration} et~al.,}{{Astropy
  Collaboration} et~al.}{2018}]{Astropy2018}
{Astropy Collaboration} et~al., 2018, \mn@doi [\aj] {10.3847/1538-3881/aabc4f},
  \href {http://cdsads.u-strasbg.fr/abs/2018AJ....156..123A} {156, 123}

\bibitem[\protect\citeauthoryear{{Bacon} et~al.,}{{Bacon}
  et~al.}{2010}]{Bacon2010}
{Bacon} R.,  et~al., 2010, in \procspie. p. 773508, \mn@doi{10.1117/12.856027}

\bibitem[\protect\citeauthoryear{{Bailer-Jones}, {Rybizki}, {Fouesneau},
  {Mantelet}  \& {Andrae}}{{Bailer-Jones} et~al.}{2018}]{bailer2018}
{Bailer-Jones} C.~A.~L.,  {Rybizki} J.,  {Fouesneau} M.,  {Mantelet} G.,
  {Andrae} R.,  2018, \mn@doi [\aj] {10.3847/1538-3881/aacb21}, \href
  {http://adsabs.harvard.edu/abs/2018AJ....156...58B} {156, 58}

\bibitem[\protect\citeauthoryear{{Baldwin}, {Phillips}  \&
  {Terlevich}}{{Baldwin} et~al.}{1981}]{Baldwin1981}
{Baldwin} J.~A.,  {Phillips} M.~M.,   {Terlevich} R.,  1981, \mn@doi [\pasp]
  {10.1086/130766}, \href {http://adsabs.harvard.edu/abs/1981PASP...93....5B}
  {93, 5}

\bibitem[\protect\citeauthoryear{{Balick}, {Rugers}, {Terzian}  \&
  {Chengalur}}{{Balick} et~al.}{1993}]{Balick1993}
{Balick} B.,  {Rugers} M.,  {Terzian} Y.,   {Chengalur} J.~N.,  1993, \mn@doi
  [\apj] {10.1086/172881}, \href
  {http://cdsads.u-strasbg.fr/abs/1993ApJ...411..778B} {411, 778}

\bibitem[\protect\citeauthoryear{{Balick}, {Perinotto}, {Maccioni}, {Terzian}
  \& {Hajian}}{{Balick} et~al.}{1994}]{Balick1994}
{Balick} B.,  {Perinotto} M.,  {Maccioni} A.,  {Terzian} Y.,   {Hajian} A.,
  1994, \mn@doi [\apj] {10.1086/173932}, \href
  {http://cdsads.u-strasbg.fr/abs/1994ApJ...424..800B} {424, 800}

\bibitem[\protect\citeauthoryear{{Barr{\'{\i}}a} \&
  {Kimeswenger}}{{Barr{\'{\i}}a} \& {Kimeswenger}}{2018a}]{Barria2018c}
{Barr{\'{\i}}a} D.,  {Kimeswenger} S.,  2018a, \mn@doi [Galaxies]
  {10.3390/galaxies6030084}, \href
  {http://cdsads.u-strasbg.fr/abs/2018Galax...6...84B} {6, 84}

\bibitem[\protect\citeauthoryear{{Barr{\'{\i}}a} \&
  {Kimeswenger}}{{Barr{\'{\i}}a} \& {Kimeswenger}}{2018b}]{Barria2018a}
{Barr{\'{\i}}a} D.,  {Kimeswenger} S.,  2018b, \mn@doi [\mnras]
  {10.1093/mnras/sty1796}, \href
  {http://cdsads.u-strasbg.fr/abs/2018MNRAS.480.1626B} {480, 1626}

\bibitem[\protect\citeauthoryear{{Barr{\'{\i}}a}, {Kimeswenger}, {Kausch}  \&
  {Goldman}}{{Barr{\'{\i}}a} et~al.}{2018}]{Barria2018b}
{Barr{\'{\i}}a} D.,  {Kimeswenger} S.,  {Kausch} W.,   {Goldman} D.~S.,  2018,
  \mn@doi [\aap] {10.1051/0004-6361/201833981}, \href
  {http://adsabs.harvard.edu/abs/2018A%26A...620A..84B} {620, A84}

\bibitem[\protect\citeauthoryear{{Basurah}, {Ali}, {Dopita}, {Alsulami}, {Amer}
   \& {Alruhaili}}{{Basurah} et~al.}{2016}]{Basurah2016}
{Basurah} H.~M.,  {Ali} A.,  {Dopita} M.~A.,  {Alsulami} R.,  {Amer} M.~A.,
  {Alruhaili} A.,  2016, \mn@doi [\mnras] {10.1093/mnras/stw468}, \href
  {http://cdsads.u-strasbg.fr/abs/2016MNRAS.458.2694B} {458, 2694}

\bibitem[\protect\citeauthoryear{{Belfiore} et~al.,}{{Belfiore}
  et~al.}{2016}]{Belfiore2016}
{Belfiore} F.,  et~al., 2016, \mn@doi [\mnras] {10.1093/mnras/stw1234}, \href
  {http://adsabs.harvard.edu/abs/2016MNRAS.461.3111B} {461, 3111}

\bibitem[\protect\citeauthoryear{{Boffin} et~al.,}{{Boffin}
  et~al.}{2018}]{Boffin2018}
{Boffin} H.~M.~J.,  et~al., 2018, preprint, \href
  {http://adsabs.harvard.edu/abs/2018arXiv180711709B} {} (\mn@eprint {arXiv}
  {1807.11709})

\bibitem[\protect\citeauthoryear{{Bohigas}}{{Bohigas}}{2003}]{Bohigas2003}
{Bohigas} J.,  2003, \rmxaa, \href
  {http://cdsads.u-strasbg.fr/abs/2003RMxAA..39..149B} {39, 149}

\bibitem[\protect\citeauthoryear{{Bohigas}}{{Bohigas}}{2008}]{bohigas2008}
{Bohigas} J.,  2008, \mn@doi [\apj] {10.1086/524977}, \href
  {http://adsabs.harvard.edu/abs/2008ApJ...674..954B} {674, 954}

\bibitem[\protect\citeauthoryear{{Bonnet} et~al.,}{{Bonnet}
  et~al.}{2004}]{Bonnet2004}
{Bonnet} H.,  et~al., 2004, The Messenger, \href
  {https://ui.adsabs.harvard.edu/abs/2004Msngr.117...17B} {117, 17}

\bibitem[\protect\citeauthoryear{{Coelho}}{{Coelho}}{2014}]{Coelho2014}
{Coelho} P.~R.~T.,  2014, \mn@doi [\mnras] {10.1093/mnras/stu365}, \href
  {https://ui.adsabs.harvard.edu/abs/2014MNRAS.440.1027C} {440, 1027}

\bibitem[\protect\citeauthoryear{{Cuevas} et~al.,}{{Cuevas}
  et~al.}{2008}]{Cuevas2008}
{Cuevas} S.,  et~al., 2008, in \procspie. p. 70146D, \mn@doi{10.1117/12.788024}

\bibitem[\protect\citeauthoryear{{Danehkar}}{{Danehkar}}{2015}]{Danehkar2015b}
{Danehkar} A.,  2015, \mn@doi [\apj] {10.1088/0004-637X/815/1/35}, \href
  {https://ui.adsabs.harvard.edu/abs/2015ApJ...815...35D} {815, 35}

\bibitem[\protect\citeauthoryear{{Danehkar} \& {Parker}}{{Danehkar} \&
  {Parker}}{2015}]{Danehkar2015}
{Danehkar} A.,  {Parker} Q.~A.,  2015, \mn@doi [\mnras]
  {10.1093/mnrasl/slv022}, \href
  {http://cdsads.u-strasbg.fr/abs/2015MNRAS.449L..56D} {449, L56}

\bibitem[\protect\citeauthoryear{{Danehkar}, {Parker}  \&
  {Ercolano}}{{Danehkar} et~al.}{2013}]{Danehkar2013}
{Danehkar} A.,  {Parker} Q.~A.,   {Ercolano} B.,  2013, \mn@doi [\mnras]
  {10.1093/mnras/stt1116}, \href
  {https://ui.adsabs.harvard.edu/abs/2013MNRAS.434.1513D} {434, 1513}

\bibitem[\protect\citeauthoryear{{Danehkar}, {Todt}, {Ercolano}  \&
  {Kniazev}}{{Danehkar} et~al.}{2014}]{Danehkar2014}
{Danehkar} A.,  {Todt} H.,  {Ercolano} B.,   {Kniazev} A.~Y.,  2014, \mn@doi
  [\mnras] {10.1093/mnras/stu203}, \href
  {https://ui.adsabs.harvard.edu/abs/2014MNRAS.439.3605D} {439, 3605}

\bibitem[\protect\citeauthoryear{{Danehkar}, {Parker}  \& {Steffen}}{{Danehkar}
  et~al.}{2016}]{Danehkar2016}
{Danehkar} A.,  {Parker} Q.~A.,   {Steffen} W.,  2016, \mn@doi [\aj]
  {10.3847/0004-6256/151/2/38}, \href
  {https://ui.adsabs.harvard.edu/abs/2016AJ....151...38D} {151, 38}

\bibitem[\protect\citeauthoryear{{Delgado-Inglada}, {Morisset}  \&
  {Stasi{\'n}ska}}{{Delgado-Inglada} et~al.}{2014}]{Inglada2014}
{Delgado-Inglada} G.,  {Morisset} C.,   {Stasi{\'n}ska} G.,  2014, \mn@doi
  [\mnras] {10.1093/mnras/stu341}, \href
  {http://adsabs.harvard.edu/abs/2014MNRAS.440..536D} {440, 536}

\bibitem[\protect\citeauthoryear{{Derlopa}, {Akras}, {Boumis}  \&
  {Steffen}}{{Derlopa} et~al.}{2019}]{Derlopa2019}
{Derlopa} S.,  {Akras} S.,  {Boumis} P.,   {Steffen} W.,  2019, \mn@doi
  [\mnras] {10.1093/mnras/stz193}, \href
  {http://cdsads.u-strasbg.fr/abs/2019MNRAS.484.3746D} {484, 3746}

\bibitem[\protect\citeauthoryear{{Dopita}, {Hart}, {McGregor}, {Oates},
  {Bloxham}  \& {Jones}}{{Dopita} et~al.}{2007}]{Dopita2007}
{Dopita} M.,  {Hart} J.,  {McGregor} P.,  {Oates} P.,  {Bloxham} G.,   {Jones}
  D.,  2007, \mn@doi [\apss] {10.1007/s10509-007-9510-z}, \href
  {https://ui.adsabs.harvard.edu/abs/2007Ap&SS.310..255D} {310, 255}

\bibitem[\protect\citeauthoryear{{Dopita}, {Ali}, {Sutherland}, {Nicholls}  \&
  {Amer}}{{Dopita} et~al.}{2017}]{Dopita2017}
{Dopita} M.~A.,  {Ali} A.,  {Sutherland} R.~S.,  {Nicholls} D.~C.,   {Amer}
  M.~A.,  2017, \mn@doi [\mnras] {10.1093/mnras/stx1166}, \href
  {https://ui.adsabs.harvard.edu/abs/2017MNRAS.470..839D} {470, 839}

\bibitem[\protect\citeauthoryear{{Dopita}, {Ali}, {Karakas}, {Goldman}, {Amer}
  \& {Sutherland}}{{Dopita} et~al.}{2018}]{Dopita2018}
{Dopita} M.~A.,  {Ali} A.,  {Karakas} A.~I.,  {Goldman} D.,  {Amer} M.~A.,
  {Sutherland} R.~S.,  2018, \mn@doi [\mnras] {10.1093/mnras/stx3180}, \href
  {https://ui.adsabs.harvard.edu/\#abs/2018MNRAS.475..424D} {475, 424}

\bibitem[\protect\citeauthoryear{{Eisenhauer} et~al.,}{{Eisenhauer}
  et~al.}{2003}]{Eisenhauer2003}
{Eisenhauer} F.,  et~al., 2003, in {Iye} M.,  {Moorwood} A. F.~M.,  eds,
  Society of Photo-Optical Instrumentation Engineers (SPIE) Conference Series
  Vol. 4841, \procspie. pp 1548--1561 (\mn@eprint {arXiv} {astro-ph/0306191}),
  \mn@doi{10.1117/12.459468}

\bibitem[\protect\citeauthoryear{{Ercolano}, {Barlow}, {Storey}  \&
  {Liu}}{{Ercolano} et~al.}{2003}]{Ercolano2003}
{Ercolano} B.,  {Barlow} M.~J.,  {Storey} P.~J.,   {Liu} X.~W.,  2003, \mn@doi
  [\mnras] {10.1046/j.1365-8711.2003.06371.x}, \href
  {https://ui.adsabs.harvard.edu/abs/2003MNRAS.340.1136E} {340, 1136}

\bibitem[\protect\citeauthoryear{{Ercolano}, {Barlow}  \& {Storey}}{{Ercolano}
  et~al.}{2005}]{Ercolano2005}
{Ercolano} B.,  {Barlow} M.~J.,   {Storey} P.~J.,  2005, \mn@doi [\mnras]
  {10.1111/j.1365-2966.2005.09381.x}, \href
  {https://ui.adsabs.harvard.edu/abs/2005MNRAS.362.1038E} {362, 1038}

\bibitem[\protect\citeauthoryear{{Ferland} et~al.,}{{Ferland}
  et~al.}{2017}]{Ferland2017}
{Ferland} G.~J.,  et~al., 2017, \rmxaa, \href
  {https://ui.adsabs.harvard.edu/abs/2017RMxAA..53..385F} {53, 385}

\bibitem[\protect\citeauthoryear{{Fitzpatrick}}{{Fitzpatrick}}{1999}]{Fitzpatrick1999}
{Fitzpatrick} E.~L.,  1999, \mn@doi [\pasp] {10.1086/316293}, \href
  {https://ui.adsabs.harvard.edu/abs/1999PASP..111...63F} {111, 63}

\bibitem[\protect\citeauthoryear{{Frew} \& {Parker}}{{Frew} \&
  {Parker}}{2010}]{Frew2010}
{Frew} D.~J.,  {Parker} Q.~A.,  2010, \mn@doi [\pasa] {10.1071/AS09040}, \href
  {https://ui.adsabs.harvard.edu/abs/2010PASA...27..129F} {27, 129}

\bibitem[\protect\citeauthoryear{{Gaia Collaboration}, {Brown}, {Vallenari},
  {Prusti}, {de Bruijne}, {Babusiaux}  \& {Bailer-Jones}}{{Gaia Collaboration}
  et~al.}{2018}]{gaia2018}
{Gaia Collaboration} {Brown} A.~G.~A.,  {Vallenari} A.,  {Prusti} T.,  {de
  Bruijne} J.~H.~J.,  {Babusiaux} C.,   {Bailer-Jones} C.~A.~L.,  2018,
  preprint, \href {http://adsabs.harvard.edu/abs/2018arXiv180409365G} {}
  (\mn@eprint {arXiv} {1804.09365})

\bibitem[\protect\citeauthoryear{{Gil de Paz} et~al.,}{{Gil de Paz}
  et~al.}{2016}]{GildePaz2016}
{Gil de Paz} A.,  et~al., 2016, in \procspie. p. 99081K,
  \mn@doi{10.1117/12.2231988}

\bibitem[\protect\citeauthoryear{{G{\'o}mez-Gordillo}, {Akras},
  {Gon{\c{c}}alves}  \& {Steffen}}{{G{\'o}mez-Gordillo}
  et~al.}{2020}]{Gordillo2020}
{G{\'o}mez-Gordillo} S.,  {Akras} S.,  {Gon{\c{c}}alves} D.~R.,   {Steffen} W.,
   2020, \mn@doi [\mnras] {10.1093/mnras/staa060}, \href
  {https://ui.adsabs.harvard.edu/abs/2020MNRAS.492.4097G} {492, 4097}

\bibitem[\protect\citeauthoryear{{Gon{\c{c}}alves}, {Wesson}, {Morisset},
  {Barlow}  \& {Ercolano}}{{Gon{\c{c}}alves} et~al.}{2012}]{Goncalves2012}
{Gon{\c{c}}alves} D.~R.,  {Wesson} R.,  {Morisset} C.,  {Barlow} M.,
  {Ercolano} B.,  2012, in IAU Symposium. pp 144--147 (\mn@eprint {arXiv}
  {1110.2709}), \mn@doi{10.1017/S174392131201085X}

\bibitem[\protect\citeauthoryear{{Graczyk} et~al.,}{{Graczyk}
  et~al.}{2019}]{Graczyk2019}
{Graczyk} D.,  et~al., 2019, \mn@doi [\apj] {10.3847/1538-4357/aafbed}, \href
  {http://cdsads.u-strasbg.fr/abs/2019ApJ...872...85G} {872, 85}

\bibitem[\protect\citeauthoryear{{Grandmont}, {Drissen}, {Mandar}, {Thibault}
  \& {Baril}}{{Grandmont} et~al.}{2012}]{Grandmont2012}
{Grandmont} F.,  {Drissen} L.,  {Mandar} J.,  {Thibault} S.,   {Baril} M.,
  2012, in \procspie. p. 84460U, \mn@doi{10.1117/12.926782}

\bibitem[\protect\citeauthoryear{{Groenewegen}}{{Groenewegen}}{2018}]{Groenewegen2018}
{Groenewegen} M.~A.~T.,  2018, \mn@doi [\aap] {10.1051/0004-6361/201833478},
  \href {https://ui.adsabs.harvard.edu/abs/2018A&A...619A...8G} {619, A8}

\bibitem[\protect\citeauthoryear{{Guerrero}, {Miranda}, {Ramos-Larios}  \&
  {V{\'a}zquez}}{{Guerrero} et~al.}{2013}]{Guerrero2013}
{Guerrero} M.~A.,  {Miranda} L.~F.,  {Ramos-Larios} G.,   {V{\'a}zquez} R.,
  2013, \mn@doi [\aap] {10.1051/0004-6361/201220592}, \href
  {http://cdsads.u-strasbg.fr/abs/2013A%26A...551A..53G} {551, A53}

\bibitem[\protect\citeauthoryear{{Hall} et~al.,}{{Hall}
  et~al.}{2019}]{Hall2019}
{Hall} O.~J.,  et~al., 2019, \mn@doi [\mnras] {10.1093/mnras/stz1092}, \href
  {https://ui.adsabs.harvard.edu/abs/2019MNRAS.486.3569H} {486, 3569}

\bibitem[\protect\citeauthoryear{{Henry}, {Kwitter}, {Jaskot}, {Balick},
  {Morrison}  \& {Milingo}}{{Henry} et~al.}{2010}]{Henry2010}
{Henry} R.~B.~C.,  {Kwitter} K.~B.,  {Jaskot} A.~E.,  {Balick} B.,  {Morrison}
  M.~A.,   {Milingo} J.~B.,  2010, \mn@doi [\apj]
  {10.1088/0004-637X/724/1/748}, \href
  {http://adsabs.harvard.edu/abs/2010ApJ...724..748H} {724, 748}

\bibitem[\protect\citeauthoryear{{Ho} et~al.,}{{Ho} et~al.}{2014}]{Ho2014}
{Ho} I.-T.,  et~al., 2014, \mn@doi [\mnras] {10.1093/mnras/stu1653}, \href
  {http://adsabs.harvard.edu/abs/2014MNRAS.444.3894H} {444, 3894}

\bibitem[\protect\citeauthoryear{Hunter}{Hunter}{2007}]{Hunter2007}
Hunter J.~D.,  2007, \mn@doi [Computing In Science \& Engineering]
  {10.1109/MCSE.2007.55}, 9, 90

\bibitem[\protect\citeauthoryear{{Karakas}}{{Karakas}}{2010}]{Karakas2010}
{Karakas} A.~I.,  2010, \mn@doi [\mnras] {10.1111/j.1365-2966.2009.16198.x},
  \href {http://cdsads.u-strasbg.fr/abs/2010MNRAS.403.1413K} {403, 1413}

\bibitem[\protect\citeauthoryear{{Khan} et~al.,}{{Khan}
  et~al.}{2019}]{Khan2019}
{Khan} S.,  et~al., 2019, \mn@doi [\aap] {10.1051/0004-6361/201935304}, \href
  {https://ui.adsabs.harvard.edu/abs/2019A&A...628A..35K} {628, A35}

\bibitem[\protect\citeauthoryear{{Kingsburgh} \& {Barlow}}{{Kingsburgh} \&
  {Barlow}}{1994}]{Kingsburgh1994}
{Kingsburgh} R.~L.,  {Barlow} M.~J.,  1994, \mn@doi [\mnras]
  {10.1093/mnras/271.2.257}, \href
  {http://cdsads.u-strasbg.fr/abs/1994MNRAS.271..257K} {271, 257}

\bibitem[\protect\citeauthoryear{Kokoska \& Zwillinger}{Kokoska \&
  Zwillinger}{2000}]{kokoska2000}
Kokoska S.,  Zwillinger D.,  2000, CRC Standard Probability and Statistics
  Tables and Formulae, Student Edition, student edn.
Boca Raton : CRC Press, \url
  {https://ebookcentral.proquest.com/lib/qut/detail.action?docID=263071}

\bibitem[\protect\citeauthoryear{{Koleva}, {Prugniel}, {Bouchard}  \&
  {Wu}}{{Koleva} et~al.}{2009}]{Koleva2009}
{Koleva} M.,  {Prugniel} P.,  {Bouchard} A.,   {Wu} Y.,  2009, \mn@doi [\aap]
  {10.1051/0004-6361/200811467}, \href
  {http://adsabs.harvard.edu/abs/2009A%26A...501.1269K} {501, 1269}

\bibitem[\protect\citeauthoryear{{Lagrois}, {Joncas}, {Drissen}, {Martin},
  {Rousseau-Nepton}  \& {Alarie}}{{Lagrois} et~al.}{2015}]{Lagrois2015}
{Lagrois} D.,  {Joncas} G.,  {Drissen} L.,  {Martin} T.,  {Rousseau-Nepton} L.,
    {Alarie} A.,  2015, \mn@doi [\mnras] {10.1093/mnras/stv070}, \href
  {https://ui.adsabs.harvard.edu/abs/2015MNRAS.448.1584L} {448, 1584}

\bibitem[\protect\citeauthoryear{{Laha}, {Tyndall}, {Keenan}, {Ballance},
  {Ramsbottom}, {Ferland}  \& {Hibbert}}{{Laha} et~al.}{2017}]{Laha2017}
{Laha} S.,  {Tyndall} N.~B.,  {Keenan} F.~P.,  {Ballance} C.~P.,  {Ramsbottom}
  C.~A.,  {Ferland} G.~J.,   {Hibbert} A.,  2017, \mn@doi [\apj]
  {10.3847/1538-4357/aa7071}, \href
  {http://cdsads.u-strasbg.fr/abs/2017ApJ...841....3L} {841, 3}

\bibitem[\protect\citeauthoryear{{Le F{\`e}vre} et~al.,}{{Le F{\`e}vre}
  et~al.}{2003}]{Lafevre2003}
{Le F{\`e}vre} O.,  et~al., 2003, in {Iye} M.,  {Moorwood} A.~F.~M.,  eds,
  \procspie Vol. 4841, Instrument Design and Performance for Optical/Infrared
  Ground-based Telescopes. pp 1670--1681, \mn@doi{10.1117/12.460959}

\bibitem[\protect\citeauthoryear{{Leal-Ferreira}, {Gon{\c{c}}alves}, {Monteiro}
   \& {Richards}}{{Leal-Ferreira} et~al.}{2011}]{Ferreira2011}
{Leal-Ferreira} M.~L.,  {Gon{\c{c}}alves} D.~R.,  {Monteiro} H.,   {Richards}
  J.~W.,  2011, \mn@doi [\mnras] {10.1111/j.1365-2966.2010.17776.x}, \href
  {https://ui.adsabs.harvard.edu/abs/2011MNRAS.411.1395L} {411, 1395}

\bibitem[\protect\citeauthoryear{{Leonidaki}, {Boumis}  \& {Zezas}}{{Leonidaki}
  et~al.}{2013}]{leonidaki2013}
{Leonidaki} I.,  {Boumis} P.,   {Zezas} A.,  2013, \mn@doi [\mnras]
  {10.1093/mnras/sts324}, \href
  {http://cdsads.u-strasbg.fr/abs/2013MNRAS.429..189L} {429, 189}

\bibitem[\protect\citeauthoryear{{Leung} \& {Bovy}}{{Leung} \&
  {Bovy}}{2019}]{Leung2019}
{Leung} H.~W.,  {Bovy} J.,  2019, \mn@doi [\mnras] {10.1093/mnras/stz2245},
  \href {https://ui.adsabs.harvard.edu/abs/2019MNRAS.489.2079L} {489, 2079}

\bibitem[\protect\citeauthoryear{{Lindegren} et~al.,}{{Lindegren}
  et~al.}{2018}]{Lindegren2018}
{Lindegren} L.,  et~al., 2018, \mn@doi [\aap] {10.1051/0004-6361/201832727},
  \href {http://cdsads.u-strasbg.fr/abs/2018A%26A...616A...2L} {616, A2}

\bibitem[\protect\citeauthoryear{{McGregor} et~al.,}{{McGregor}
  et~al.}{2003}]{McGregor2003}
{McGregor} P.~J.,  et~al., 2003, in {Iye} M.,  {Moorwood} A. F.~M.,  eds,
  Society of Photo-Optical Instrumentation Engineers (SPIE) Conference Series
  Vol. 4841, \procspie. pp 1581--1591, \mn@doi{10.1117/12.459448}

\bibitem[\protect\citeauthoryear{{Monteiro}, {Dias}  \& {Caetano}}{{Monteiro}
  et~al.}{2010}]{Monteiro2010}
{Monteiro} H.,  {Dias} W.~S.,   {Caetano} T.~C.,  2010, \mn@doi [\aap]
  {10.1051/0004-6361/200913677}, \href
  {http://adsabs.harvard.edu/abs/2010A%26A...516A...2M} {516, A2}

\bibitem[\protect\citeauthoryear{{Monteiro}, {Gon{\c c}alves}, {Leal-Ferreira}
  \& {Corradi}}{{Monteiro} et~al.}{2013}]{Hektor2013}
{Monteiro} H.,  {Gon{\c c}alves} D.~R.,  {Leal-Ferreira} M.~L.,   {Corradi}
  R.~L.~M.,  2013, \mn@doi [\aap] {10.1051/0004-6361/201322220}, \href
  {http://cdsads.u-strasbg.fr/abs/2013A%26A...560A.102M} {560, A102}

\bibitem[\protect\citeauthoryear{{Morisset}}{{Morisset}}{2017}]{Morisset2017}
{Morisset} C.,  2017, in {Liu} X.,  {Stanghellini} L.,   {Karakas} A.,  eds,
  IAU Symposium Vol. 323, Planetary Nebulae: Multi-Wavelength Probes of Stellar
  and Galactic Evolution. pp 43--50 (\mn@eprint {arXiv} {1612.04242}),
  \mn@doi{10.1017/S1743921317001004}

\bibitem[\protect\citeauthoryear{{Morisset}}{{Morisset}}{2018}]{Morisset2018}
{Morisset} C.,  2018, in Walking the Line 2018. p.~2,
  \mn@doi{10.5281/zenodo.1206115}

\bibitem[\protect\citeauthoryear{{Morisset}, {Delgado-Inglada}  \&
  {Flores-Fajardo}}{{Morisset} et~al.}{2015}]{Morisset2015}
{Morisset} C.,  {Delgado-Inglada} G.,   {Flores-Fajardo} N.,  2015, \rmxaa,
  \href {http://adsabs.harvard.edu/abs/2015RMxAA..51..103M} {51, 103}

\bibitem[\protect\citeauthoryear{{Muna} et~al.,}{{Muna}
  et~al.}{2016}]{Muna2016}
{Muna} D.,  et~al., 2016, arXiv:1610.03159, \href
  {http://cdsads.u-strasbg.fr/abs/2016arXiv161003159M} {}

\bibitem[\protect\citeauthoryear{{Muraveva}, {Delgado}, {Clementini}, {Sarro}
  \& {Garofalo}}{{Muraveva} et~al.}{2018}]{Muraveva2018}
{Muraveva} T.,  {Delgado} H.~E.,  {Clementini} G.,  {Sarro} L.~M.,   {Garofalo}
  A.,  2018, \mn@doi [\mnras] {10.1093/mnras/sty2241}, \href
  {https://ui.adsabs.harvard.edu/abs/2018MNRAS.481.1195M} {481, 1195}

\bibitem[\protect\citeauthoryear{{{\"O}ttl}, {Kimeswenger}  \&
  {Zijlstra}}{{{\"O}ttl} et~al.}{2014}]{ottl2014}
{{\"O}ttl} S.,  {Kimeswenger} S.,   {Zijlstra} A.~A.,  2014, \mn@doi [\aap]
  {10.1051/0004-6361/201323205}, \href
  {https://ui.adsabs.harvard.edu/abs/2014A&A...565A..87O} {565, A87}

\bibitem[\protect\citeauthoryear{{Peimbert}}{{Peimbert}}{1978}]{Peimbert1978}
{Peimbert} M.,  1978, in {Terzian} Y.,  ed.,  IAU Symposium Vol. 76, Planetary
  Nebulae. pp 215--224

\bibitem[\protect\citeauthoryear{{Perinotto}, {Bencini}, {Pasquali},
  {Manchado}, {Rodriguez Espinosa}  \& {Stanga}}{{Perinotto}
  et~al.}{1999}]{Perinotto1999}
{Perinotto} M.,  {Bencini} C.~G.,  {Pasquali} A.,  {Manchado} A.,  {Rodriguez
  Espinosa} J.~M.,   {Stanga} R.,  1999, \aap, \href
  {http://cdsads.u-strasbg.fr/abs/1999A%26A...347..967P} {347, 967}

\bibitem[\protect\citeauthoryear{{Phillips} \& {Guzman}}{{Phillips} \&
  {Guzman}}{1998}]{phillps1998}
{Phillips} J.~P.,  {Guzman} V.,  1998, \mn@doi [\aaps] {10.1051/aas:1998366},
  \href {http://cdsads.u-strasbg.fr/abs/1998A%26AS..130..465P} {130, 465}

\bibitem[\protect\citeauthoryear{{Pottasch} \& {Bernard-Salas}}{{Pottasch} \&
  {Bernard-Salas}}{2010}]{Pottasch2010}
{Pottasch} S.~R.,  {Bernard-Salas} J.,  2010, \mn@doi [\aap]
  {10.1051/0004-6361/201014009}, \href
  {http://cdsads.u-strasbg.fr/abs/2010A%26A...517A..95P} {517, A95}

\bibitem[\protect\citeauthoryear{{Prugniel} \& {Soubiran}}{{Prugniel} \&
  {Soubiran}}{2001}]{Prugniel2001}
{Prugniel} P.,  {Soubiran} C.,  2001, \mn@doi [\aap]
  {10.1051/0004-6361:20010163}, \href
  {https://ui.adsabs.harvard.edu/abs/2001A&A...369.1048P} {369, 1048}

\bibitem[\protect\citeauthoryear{{Prugniel}, {Soubiran}, {Koleva}  \& {Le
  Borgne}}{{Prugniel} et~al.}{2007}]{Prugniel2007}
{Prugniel} P.,  {Soubiran} C.,  {Koleva} M.,   {Le Borgne} D.,  2007, VizieR
  Online Data Catalog, \href
  {http://adsabs.harvard.edu/abs/2007yCat.3251....0P} {3251}

\bibitem[\protect\citeauthoryear{{Ramos-Larios} et~al.,}{{Ramos-Larios}
  et~al.}{2018}]{Ramos2018}
{Ramos-Larios} G.,  et~al., 2018, \mn@doi [\mnras] {10.1093/mnras/stx3256},
  \href {http://cdsads.u-strasbg.fr/abs/2018MNRAS.475..932R} {475, 932}

\bibitem[\protect\citeauthoryear{{Riesgo} \& {L{\'o}pez}}{{Riesgo} \&
  {L{\'o}pez}}{2006}]{Riesgo2006}
{Riesgo} H.,  {L{\'o}pez} J.~A.,  2006, \rmxaa, \href
  {http://adsabs.harvard.edu/abs/2006RMxAA..42...47R} {42, 47}

\bibitem[\protect\citeauthoryear{{Riess} et~al.,}{{Riess}
  et~al.}{2018}]{Riess2018}
{Riess} A.~G.,  et~al., 2018, \mn@doi [\apj] {10.3847/1538-4357/aac82e}, \href
  {https://ui.adsabs.harvard.edu/abs/2018ApJ...861..126R} {861, 126}

\bibitem[\protect\citeauthoryear{{Sabbadin}, {Minello}  \&
  {Bianchini}}{{Sabbadin} et~al.}{1977}]{SMB1977}
{Sabbadin} F.,  {Minello} S.,   {Bianchini} A.,  1977, \aap, \href
  {http://adsabs.harvard.edu/abs/1977A%26A....60..147S} {60, 147}

\bibitem[\protect\citeauthoryear{{Sabin} et~al.,}{{Sabin}
  et~al.}{2013}]{sabin2013}
{Sabin} L.,  et~al., 2013, \mn@doi [\mnras] {10.1093/mnras/stt160}, \href
  {http://adsabs.harvard.edu/abs/2013MNRAS.431..279S} {431, 279}

\bibitem[\protect\citeauthoryear{{S{\'a}nchez}}{{S{\'a}nchez}}{2013}]{sanchez2013}
{S{\'a}nchez} S.~F.,  2013, \mn@doi [Advances in Astronomy]
  {10.1155/2013/596501}, \href
  {http://adsabs.harvard.edu/abs/2013AdAst2013E..25S} {2013, 1}

\bibitem[\protect\citeauthoryear{{S{\'a}nchez} et~al.,}{{S{\'a}nchez}
  et~al.}{2012}]{sanchez2012}
{S{\'a}nchez} S.~F.,  et~al., 2012, \mn@doi [\aap]
  {10.1051/0004-6361/201117353}, \href
  {http://adsabs.harvard.edu/abs/2012A%26A...538A...8S} {538, A8}

\bibitem[\protect\citeauthoryear{{Sch{\"o}nrich}, {McMillan}  \&
  {Eyer}}{{Sch{\"o}nrich} et~al.}{2019}]{Schonrich2019}
{Sch{\"o}nrich} R.,  {McMillan} P.,   {Eyer} L.,  2019, \mn@doi [\mnras]
  {10.1093/mnras/stz1451}, \href
  {https://ui.adsabs.harvard.edu/abs/2019MNRAS.487.3568S} {487, 3568}

\bibitem[\protect\citeauthoryear{{Singh} et~al.,}{{Singh}
  et~al.}{2013}]{singh2013}
{Singh} R.,  et~al., 2013, \mn@doi [\aap] {10.1051/0004-6361/201322062}, \href
  {http://adsabs.harvard.edu/abs/2013A%26A...558A..43S} {558, A43}

\bibitem[\protect\citeauthoryear{{Tsamis}, {Walsh}, {P{\'e}quignot}, {Barlow},
  {Danziger}  \& {Liu}}{{Tsamis} et~al.}{2008}]{Tsamis2008}
{Tsamis} Y.~G.,  {Walsh} J.~R.,  {P{\'e}quignot} D.,  {Barlow} M.~J.,
  {Danziger} I.~J.,   {Liu} X.-W.,  2008, \mn@doi [\mnras]
  {10.1111/j.1365-2966.2008.13051.x}, \href
  {http://cdsads.u-strasbg.fr/abs/2008MNRAS.386...22T} {386, 22}

\bibitem[\protect\citeauthoryear{Upton \& Cook}{Upton \&
  Cook}{1997}]{upton1997}
Upton G.,  Cook I.,  1997, Understanding statistics.
Oxford : Oxford University Press

\bibitem[\protect\citeauthoryear{{Veilleux} \& {Osterbrock}}{{Veilleux} \&
  {Osterbrock}}{1987}]{Veilleux1987}
{Veilleux} S.,  {Osterbrock} D.~E.,  1987, \mn@doi [\apjs] {10.1086/191166},
  \href {http://adsabs.harvard.edu/abs/1987ApJS...63..295V} {63, 295}

\bibitem[\protect\citeauthoryear{{Virtanen} et~al.,}{{Virtanen}
  et~al.}{2020}]{SciPy2020}
{Virtanen} P.,  et~al., 2020, \mn@doi [Nature Methods]
  {https://doi.org/10.1038/s41592-019-0686-2}, \href {https://rdcu.be/b08Wh} {}

\bibitem[\protect\citeauthoryear{{Walsh}, {Monreal-Ibero}, {Barlow}, {Ueta},
  {Wesson}  \& {Zijlstra}}{{Walsh} et~al.}{2016}]{Walsh2016}
{Walsh} J.~R.,  {Monreal-Ibero} A.,  {Barlow} M.~J.,  {Ueta} T.,  {Wesson} R.,
   {Zijlstra} A.~A.,  2016, \mn@doi [\aap] {10.1051/0004-6361/201527988}, \href
  {http://cdsads.u-strasbg.fr/abs/2016A%26A...588A.106W} {588, A106}

\bibitem[\protect\citeauthoryear{{Walsh} et~al.,}{{Walsh}
  et~al.}{2018}]{Walsh2018}
{Walsh} J.~R.,  et~al., 2018, \mn@doi [\aap] {10.1051/0004-6361/201833445},
  \href {https://ui.adsabs.harvard.edu/abs/2018A&A...620A.169W} {620, A169}

\bibitem[\protect\citeauthoryear{{Weidmann} \& {Gamen}}{{Weidmann} \&
  {Gamen}}{2011}]{Weidmann2011}
{Weidmann} W.~A.,  {Gamen} R.,  2011, \mn@doi [\aap]
  {10.1051/0004-6361/200913984}, \href
  {http://adsabs.harvard.edu/abs/2011A%26A...526A...6W} {526, A6}

\bibitem[\protect\citeauthoryear{{Wesson}}{{Wesson}}{2016}]{Wesson16}
{Wesson} R.,  2016, \mn@doi [\mnras] {10.1093/mnras/stv2946}, \href
  {http://adsabs.harvard.edu/abs/2016MNRAS.456.3774W} {456, 3774}

\bibitem[\protect\citeauthoryear{{Wesson}, {Stock}  \& {Scicluna}}{{Wesson}
  et~al.}{2012}]{Wesson2012}
{Wesson} R.,  {Stock} D.~J.,   {Scicluna} P.,  2012, \mn@doi [\mnras]
  {10.1111/j.1365-2966.2012.20863.x}, \href
  {https://ui.adsabs.harvard.edu/abs/2012MNRAS.422.3516W} {422, 3516}

\bibitem[\protect\citeauthoryear{{Xu}, {Zhang}, {Reid}, {Zheng}  \&
  {Wang}}{{Xu} et~al.}{2019}]{Xu2019}
{Xu} S.,  {Zhang} B.,  {Reid} M.~J.,  {Zheng} X.,   {Wang} G.,  2019, \mn@doi
  [\apj] {10.3847/1538-4357/ab0e83}, \href
  {https://ui.adsabs.harvard.edu/abs/2019ApJ...875..114X} {875, 114}

\bibitem[\protect\citeauthoryear{{Zhang} \& {Liu}}{{Zhang} \&
  {Liu}}{2002}]{Zhang2002}
{Zhang} Y.,  {Liu} X.-W.,  2002, \mn@doi [\mnras]
  {10.1046/j.1365-8711.2002.05929.x}, \href
  {http://cdsads.u-strasbg.fr/abs/2002MNRAS.337..499Z} {337, 499}

\bibitem[\protect\citeauthoryear{{Zhang}, {Fang}, {Chau}, {Hsia}, {Liu}, {Kwok}
   \& {Koning}}{{Zhang} et~al.}{2012}]{Zhang2012}
{Zhang} Y.,  {Fang} X.,  {Chau} W.,  {Hsia} C.-H.,  {Liu} X.-W.,  {Kwok} S.,
  {Koning} N.,  2012, \mn@doi [\apj] {10.1088/0004-637X/754/1/28}, \href
  {http://cdsads.u-strasbg.fr/abs/2012ApJ...754...28Z} {754, 28}

\bibitem[\protect\citeauthoryear{{Zinn}, {Pinsonneault}, {Huber}  \&
  {Stello}}{{Zinn} et~al.}{2019}]{Zinn2019}
{Zinn} J.~C.,  {Pinsonneault} M.~H.,  {Huber} D.,   {Stello} D.,  2019, \mn@doi
  [\apj] {10.3847/1538-4357/ab1f66}, \href
  {https://ui.adsabs.harvard.edu/abs/2019ApJ...878..136Z} {878, 136}

\bibitem[\protect\citeauthoryear{{van der Walt}, {Colbert}  \&
  {Varoquaux}}{{van der Walt} et~al.}{2011}]{Walt2011}
{van der Walt} S.,  {Colbert} S.~C.,   {Varoquaux} G.,  2011, \mn@doi
  [Computing in Science Engineering] {10.1109/MCSE.2011.37}, 13, 22

\makeatother
\end{thebibliography}



\bsp	
\label{lastpage}
\end{document}